\documentclass[aps,twocolumn,nofootinbib,preprintnumbers,
]{revtex4-1}

\usepackage[
	  pagebackref=false,
	  colorlinks=true,
      linkcolor=blue,
      urlcolor=blue,
      filecolor=black,
      citecolor=red,
      pdfstartview=FitV,
      pdftitle={},
        pdfauthor={},
        pdfsubject={},
        pdfkeywords={},
        pdfpagemode=None,
        bookmarksopen=true
      ]{hyperref}

\usepackage[normalem]{ulem}
\usepackage{amsmath}
\usepackage{enumerate}
\usepackage{amsfonts}
\usepackage{epsfig}
\usepackage{mathbbol}

\newcommand{\f}{\begin{equation}}
\newcommand{\ff}{\end{equation}}
\newcommand{\fa}{\begin{eqnarray}}
\newcommand{\ffa}{\end{eqnarray}}

\usepackage{color}

\newcommand\half{{\ensuremath{\frac{1}{2}}}}
\newcommand\p{\ensuremath{\partial}}

\newcommand{\be}{\begin{equation}}
\newcommand{\ee}{\end{equation}}
\newcommand{\bea}{\begin{eqnarray}}
\newcommand{\eea}{\end{eqnarray}}
\newcommand{\bega}{\begin{gather}}
\newcommand{\eega}{\end{gather}}

\newcommand{\bi}{\begin{itemize}}
\newcommand{\ei}{\end{itemize}}
\newcommand{\ben}{\begin{enumerate}}
\newcommand{\een}{\end{enumerate}}
\newcommand{\bca}{\begin{cases}}
\newcommand{\eca}{\end{cases}}
\newcommand{\bln}{\begin{align}}
\newcommand{\eln}{\end{align}}
\newcommand{\bst}{\begin{split}}
\newcommand{\est}{\end{split}}
\def\ie{\begin{equation}\begin{aligned}}
\def\fe{\end{aligned}\end{equation}}
\newcommand{\bma}{\le(\begin{matrix}}
\newcommand{\ema}{\end{matrix}\ri)}

\newcommand\al{{\alpha}}
\newcommand\ep{\epsilon}

\newcommand\lam{\lambda}

\newcommand\om{\omega}

\newcommand\ga{{\ensuremath{{\gamma}}}}

\newcommand\de{{\ensuremath{{\delta}}}}

\newcommand\ka{\kappa}

\newcommand\nab{{\nabla}}

\def\th{{\theta}}

\newcommand\ov{\over}
\newcommand\ha{{\half}}

\def\le{\left}
\def\ri{\right}

\newcommand\vx{{\vec x}}

\begin{document}

\preprint{MIT-CTP/5077}

\title{A dynamical phase transition from non-equilibrium dynamics of dark solitons}
\author{Minyong Guo $^{1,2}$}
\email{minyongguo@mail.bnu.edu.cn}

\author{Esko Keski-Vakkuri$^3$}
\email{esko.keski-vakkuri@helsinki.fi}

\author{Hong Liu$^4$}
\email{hong$_$liu@mit.edu}

\author{Yu Tian $^{5,6}$}
\email{ytian@ucas.ac.cn}
\author{Hongbao Zhang $^{1,7}$}
\email{hzhang@vub.ac.be}
\affiliation{$^1$Department of Physics, Beijing Normal University, Beijing 100875, China\\
 $^2$Perimeter Institute for Theoretical Physics, Waterloo, Ontario N2L 2Y5, Canada\\
$^3$ Helsinki Institute of Physics and Department of Physics, University of Helsinki, FIN 00014, Finland\\
$^4$Center for Theoretical Physics, Massachusetts Institute of Technology, Cambridge, MA 02139, USA\\
$^5$ School of Physics, University of Chinese Academy of Sciences,
Beijing 100049, China\\
$^6$ Institute of Theoretical Physics, Chinese Academy of Sciences, Beijing 100190, China\\
$^7$  Theoretische Natuurkunde, Vrije Universiteit Brussel,
and The International Solvay Institutes,\\
Pleinlaan 2, B-1050 Brussels, Belgium
}
\begin{abstract}

 We identify a novel dynamical phase transition
which results from temperature dependence of non-equilibrium dynamics of dark solitons in a superfluid. 
For a non-equilibrium superfluid system with an initial density of dark solitons, there exists a critical temperature $T_d$, 
above which the system relaxes to equilibrium by producing sound waves, while below which it goes through an intermediate phase with a finite density of vortex-antivortex pairs. In particular, as $T_d$ is approached from below, the density of vortex pairs scales as $(T_d - T)^\ga$ with critical exponent $\ga = \ha$. Such a dynamical phase transition is found to occur in both  
 the dissipative Gross-Pitaevskii equation and holographic superfluids, strong suggesting it is a universal phenomenon.

\end{abstract}
\pacs{11.25.Tq, 04.70.Bw}
\maketitle

\noindent {\it Introduction} 

\begin{figure}
\begin{center}
\includegraphics[width=9.5cm]{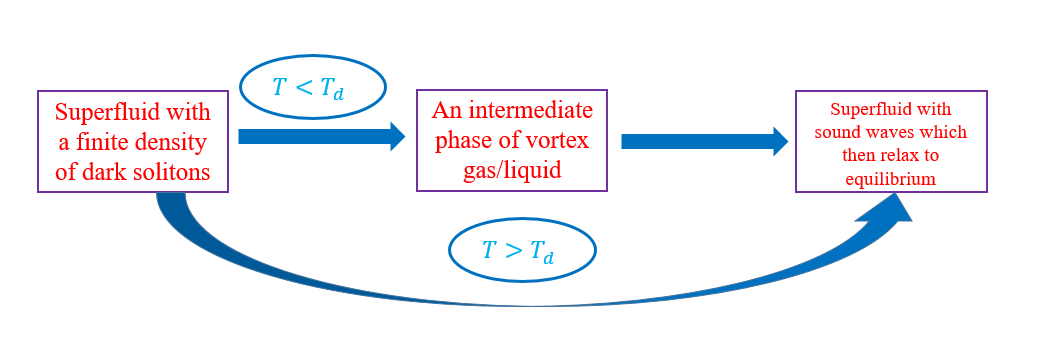}
\end{center}
\caption{For $T<T_d$ there is an intermediate phase of vortex gas/liquid. 
}
\label{fig:phases}
\end{figure}

In this paper we discuss a novel dynamical phase transition for superfluids. 
Non-equilibrium superfluid states obtained from external driving or phase/density imprinting
often contain a large number of dark solitons~\cite{Burger,Denschlag,Becker,Stellmer,Weller,Frantz,KMYZ}, which are regions of low densities in a superfluid.  The subsequent evolution of such a system
and its relaxation to equilibrium are of great interest. We argue that the dynamical 
relaxation process undergoes a ``phase transition'' at some critical temperature $T_d$. 
For $T> T_d$, the solitons directly decay into sound waves which subsequently relax to equilibrium, while 
for $T < T_d$  an intermediate phase with a finite density of vortices emerges. See FIG.~\ref{fig:phases}.
The intermediate phase is unstable  but could in principle be long-lived enough for interesting 
dynamics including turbulence to happen. The density $n_v$
of vortices, which may be considered as the ``order parameter'' of the intermediate phase,  is proportional to the density of original dark solitons and increases as the temperature 
is decreased. Near $T_d$, it has critical behavior
\be \label{che}
n_v  \propto (T_d - T)^\ga, \qquad \ga = \ha  \ .
\ee
While we are dealing with a non-equilibrium process, nevertheless a sensible temperature can be defined 
which measures the  fractions of normal and superfluid components, and characterizes the dissipation of the system. 
This is also usually how temperature is determined in experimental situations (see e.g.~\cite{B1,B2}).

The phase transition is deduced by examining in detail 
the temperature dependence of the decay of dark solitons in two classes of theories, which encompasses 
most systems currently under study in the literature. 
One class is the dissipative Gross-Pitaevskii equation (GPE)~\cite{Gross,Pitaevskii,DGP}, which is a phenomenological theory with a dissipative parameter $\kappa$ added ``by hand'' to the GPE. This theory is rather crude and requires significant modeling. 
The other class is holographic superfluid. 
Holographic duality equates certain strongly correlated systems of quantum matter without gravity to 
classical gravitational systems in a curved spacetime with one additional spatial dimension.
The gravity description provides a first-principle description of  finite temperature dissipative effects; as we will see below once the ``microscopic'' theory is fixed, all aspects of the superfluid phase are determined. There is no phenomenological modeling involved. 
Remarkably, despite that these two classes of theories are governed by completely different sets of equations, and have very different physical origins, we find qualitatively similar results.  

A dark soliton can decay by self-accelerating uniformly and then turning into sound waves, or by fragmenting into vortex pairs or filaments, a process called snake instability.  We find that for $T > T_d$ it decays dominantly by self-acceleration but by snake instabilities for $T< T_d$. We show that this pattern persists with a finite density of dark solitons, which results in the phase transition. We also present a general analytic argument for the ``critical exponent'' $\ha$ in~\eqref{che}. 
 
Instabilities of dark solitons in various superfluids were studied extensively in the literature at theoretical level (see e.g.~\cite{MLS,BA,BA1,GK,CBSDP}), mainly based on the GPE or Bogoliubov-de Gennes (BdG) equations~\cite{BdG}, which
are suitable for zero or very low temperatures. A main technical advance which enables us to find the transition is a systematic study of temperature dependence of the dynamics and decay mechanisms of dark solitons (see also~\cite{CMB,FMS,MSESL,PPBA,JPB,GNZ,Cockburn,MR,LAKT} for other finite temperature studies).


\medskip

 \noindent {\it Setup and dark soliton solutions}

The dissipative GPE for a superfluid can be written as 
\be \label{di1}
(i -\ka)  \p_t \psi  = \ \le(- {1 \ov 2} \nab^2 + |\psi|^2  - \mu \ri) \psi\ ,
\ee
where we have done some rescalings. 
The parameter $\mu$ specifies the equilibrium value of the condensate $\psi_0 = v \equiv \sqrt{\mu}$. 
For definiteness we will consider two spatial dimensions with spatial coordinates $\vec x = (x,y)$. 
$\ka > 0$ is a dissipative parameter. To see its effect, note that 
\be 
\p_t F = - 2 \ka \int d^2 \vx \, |\p_t \psi|^2, 
\ee
where $F$ is the free energy functional of the system 
\be  \label{free}
F [\psi] = \int d^2 \vx \,  \le( \ha  |\nab  \psi|^2  + \ha |\psi|^4 - \mu  |\psi|^2 \ri)  \ .
\ee
Both $\ka$ and $\mu$ should be considered as temperature dependent. 
We expect $\ka$ to increase monotonically with temperature and $\mu$ to decrease with temperature. 
In particular $\mu$ should approach $0$ as the critical 
temperature for a superfluid is approached from below. 
Below we will take $\ka$ as a proxy for temperature, i.e.  when we say increasing the temperature for the dissipative GPE, we mean increasing $\ka$.  At the moment there is no first principle to decide the precise $\ka$-dependence of $\mu$, which may be considered as a major defect of this model.

The static dark soliton solution to~\eqref{di1}, which we will denote as $\psi_S$, can be readily obtained analytically.
For definiteness we will take it to be translation invariant along $y$ direction and centered at $x=0$, i.e.,
\begin{equation}
\psi_S = \psi_S (x)=v\tanh vx \ .
\end{equation}
$\psi_S (x)$ has the following properties:
\bega \label{p1}
\psi_S^* (x) = \psi_S (x), \quad \psi_S (-x) = - \psi_S (x), \\ 
 \psi_S (x \to + \infty) \to v \; \; \text{exponentially fast} \ .
 \label{p2}
\end{gather}

Let us now turn to holographic superfluids in $2+1$ dimensions, which can be described by an Abelian-Higgs model in a $(3+1)$-dimensional anti-de Sitter (AdS$_4$) black hole spacetime~\cite{Gubser:2008px,HHH1}. The Lagrangian can be written as 
\begin{equation} \label{lag}
\mathcal{L} =-\frac{1}{4}F_{ab}F^{ab}-|(\nabla- i A) \Psi|^2-m^2|\Psi|^2
\end{equation}
with the background spacetime metric 
\begin{equation}\label{ui}
ds^2=\frac{L^2}{z^2}(-f(z)dt^2-2dtdz+dx^2 + dy^2) \ .
\end{equation}
In~\eqref{ui},  $L$ is the curvature radius of AdS and $f(z)=1-(\frac{z}{z_h})^3$ with $z=z_h$ the location of an event horizon and $z=0$ the AdS boundary. The black hole spacetime~\eqref{ui} has a Hawking temperature
$T=\frac{3}{4\pi z_h}$,
which is identified with the temperature of the dual boundary system. In~\eqref{lag} the $U(1)$ gauge field $A_a$ is dual to a conserved current $J^a$ for a $U(1)$ global symmetry in the boundary system, and the complex scalar field $\Psi$ is dual to a boundary order parameter $\psi$ charged under the $U(1)$ symmetry. The system is in a superfluid phase below some critical temperature $T_c$, when $\Psi$ develops a normalizable profile in the bulk spacetime which corresponds to $\psi$ developing a nonzero expectation value~\cite{Gubser:2008px,HHH1}. We will ignore the backreaction of $A_a$ and $\Psi$ to the background black hole geometry, an approximation which works well when the temperature is not too low~\cite{HHH2}. We are mostly interested in the regime  $T/T_c \gtrsim 0.3$ where the approximation is sufficient.

\begin{figure}
\begin{center}
\includegraphics[width=4cm]{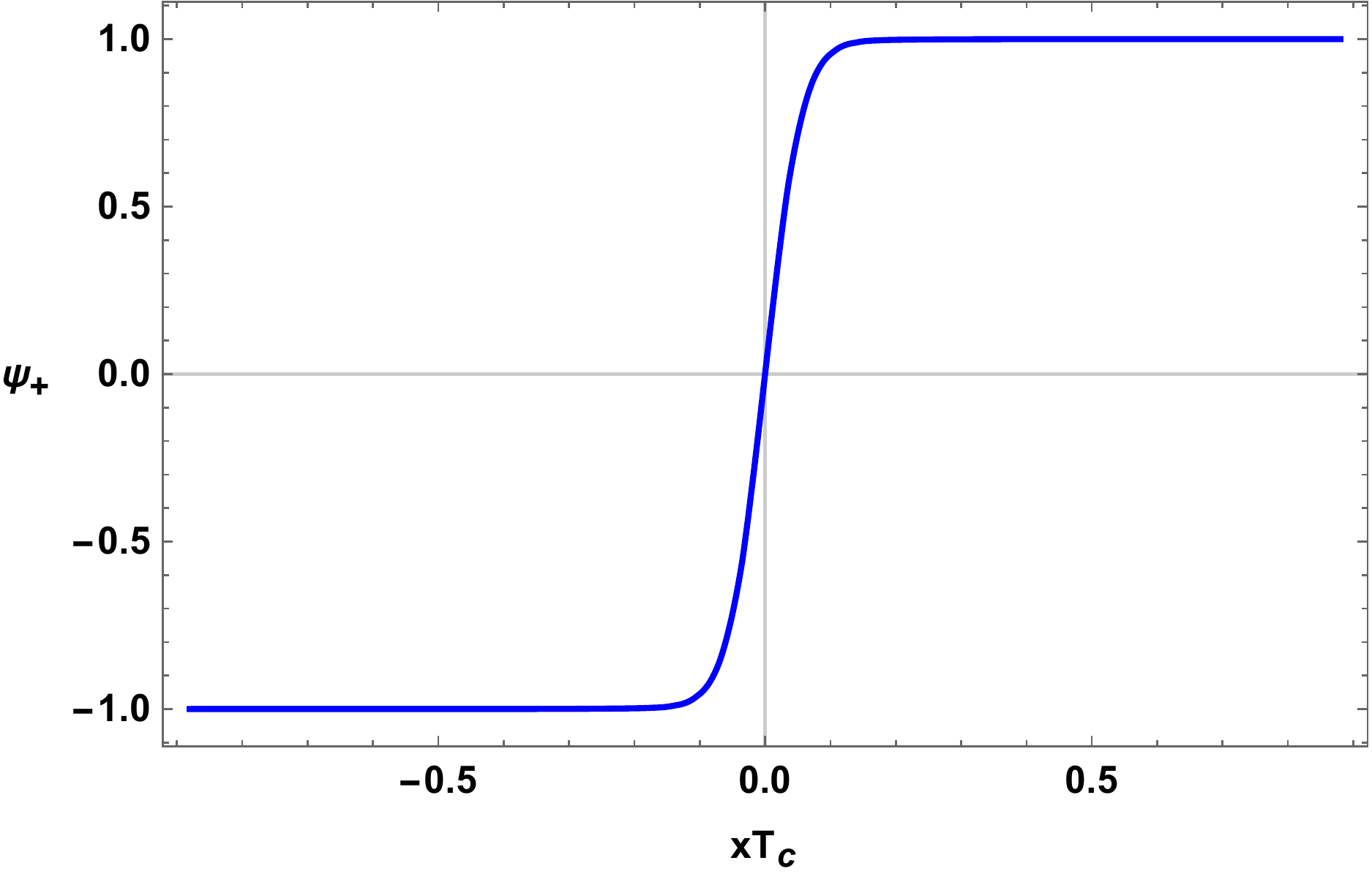}
\includegraphics[width=4cm]{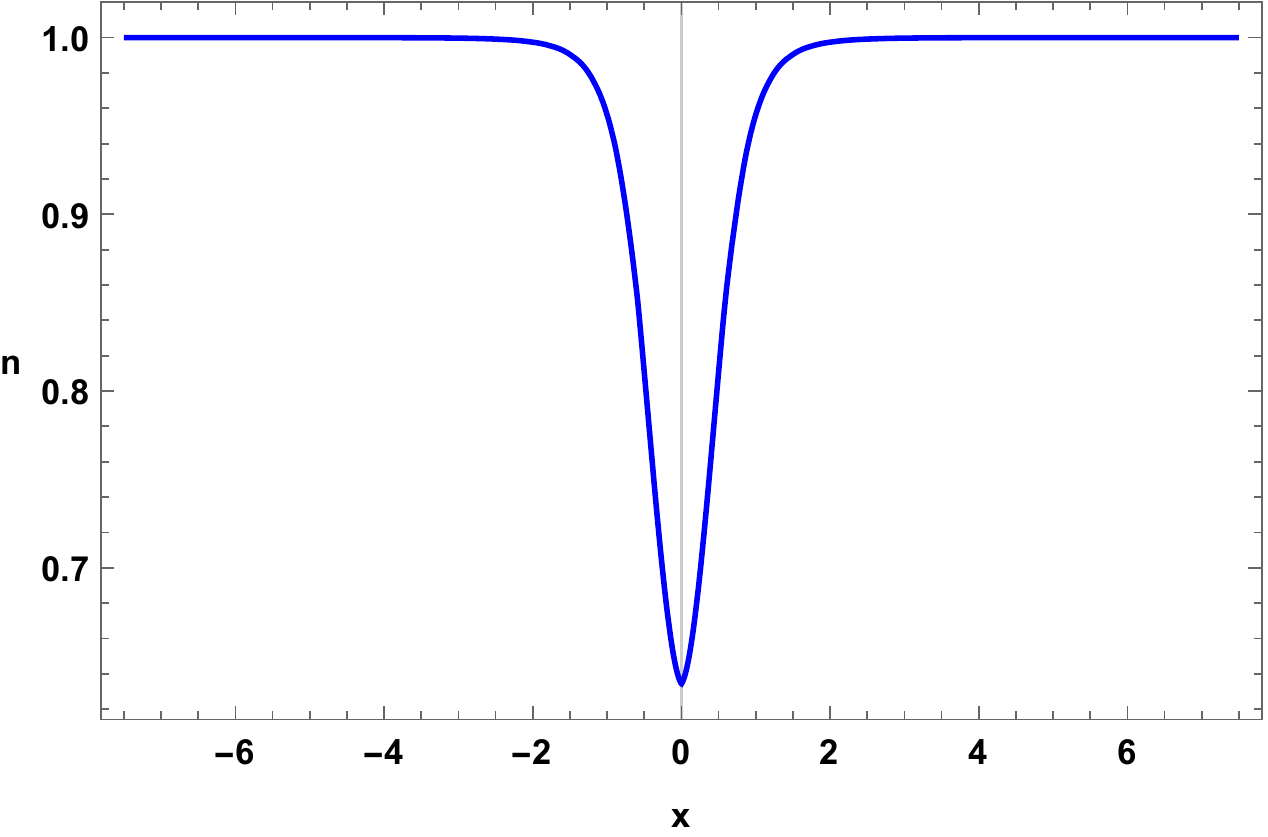}
\includegraphics[width=4cm]{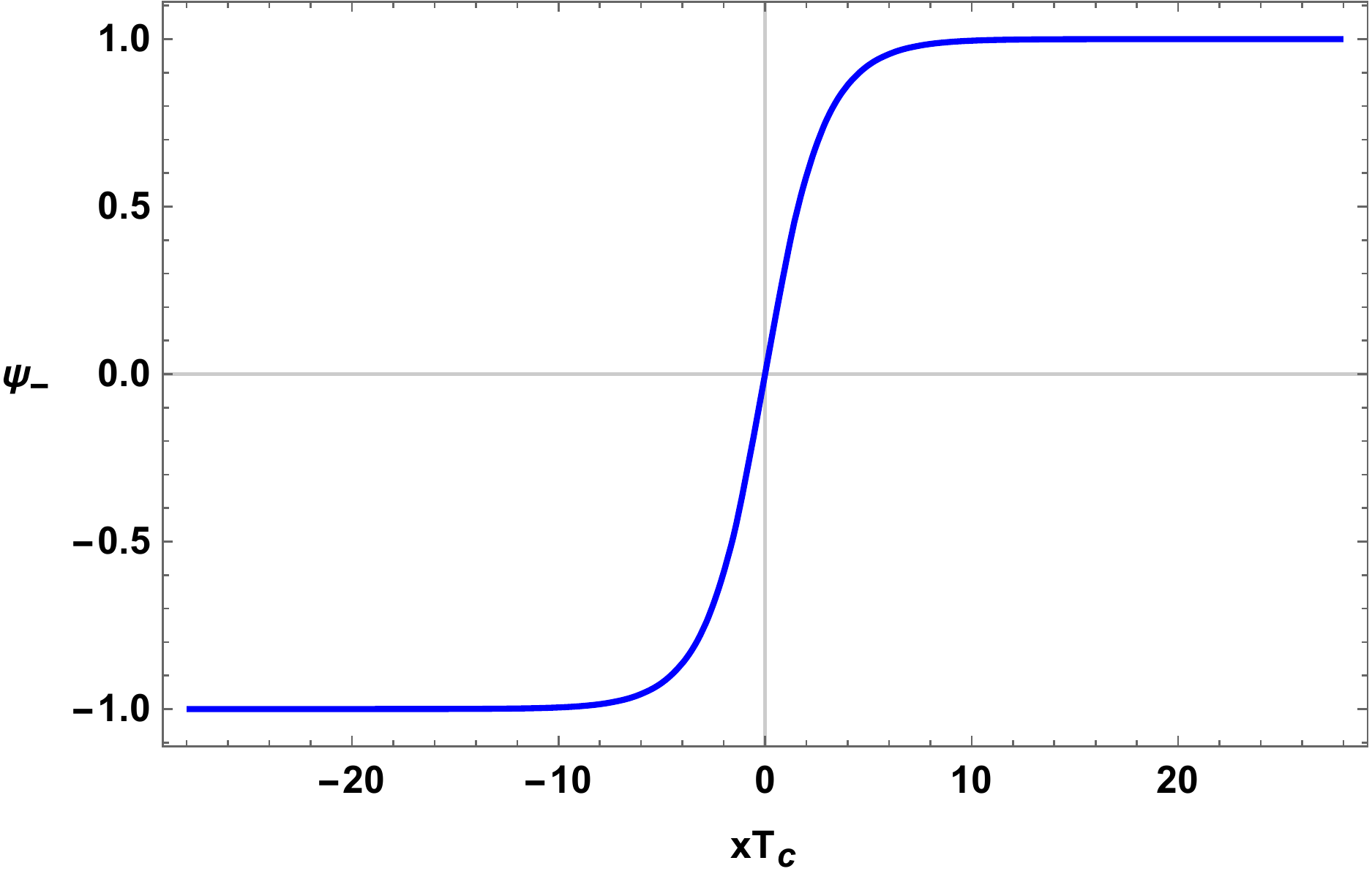}
\includegraphics[width=4cm]{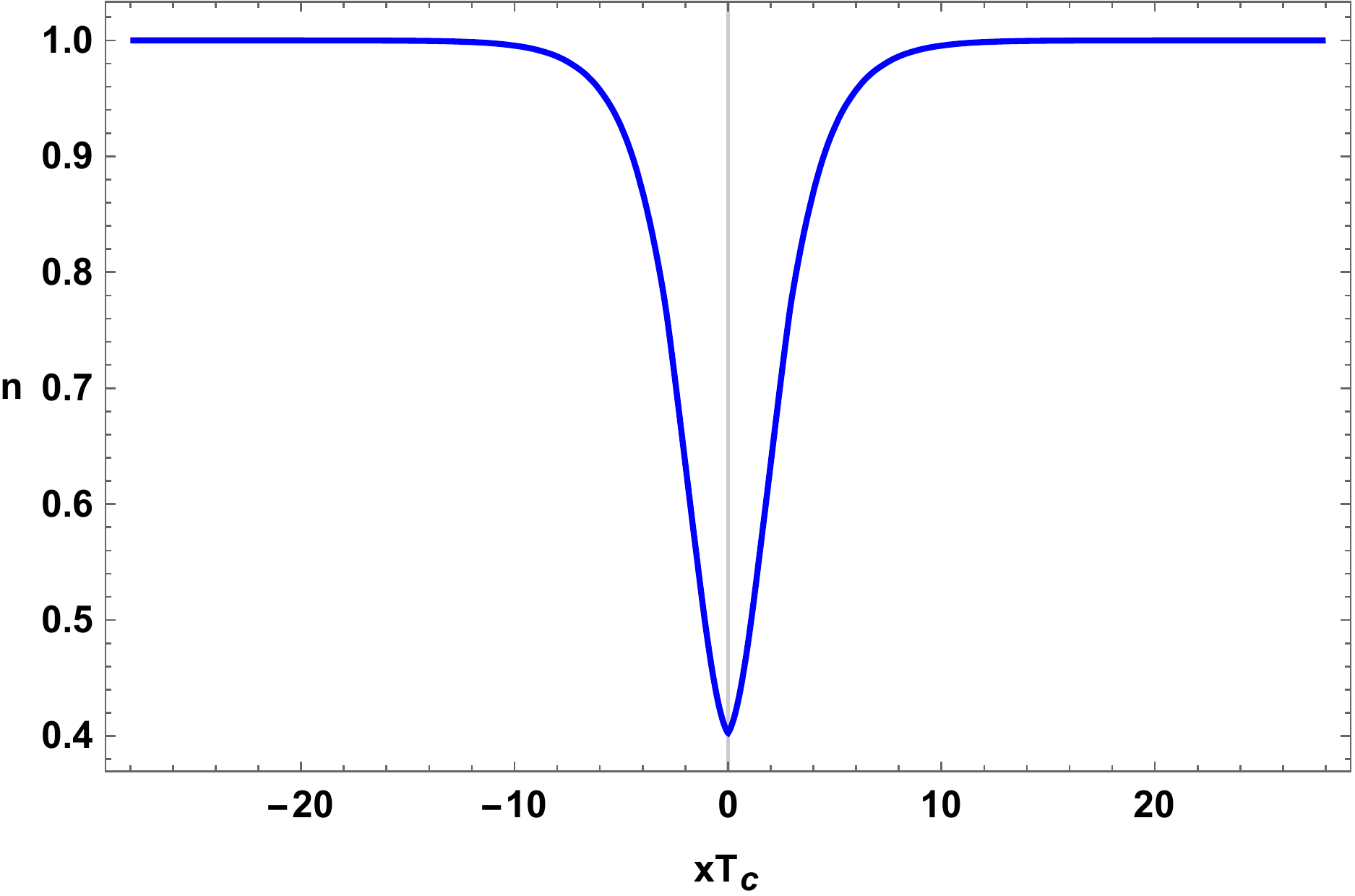}
\end{center}
\caption{Upper panel: the dark soliton solution for $\psi_+$-superfluid (BCS-like) at $\frac{T}{T_c}=0.48$, the plots are respectively profiles for the condensate and number density $n$ (normalized to $1$ at infinity). Lower panel: the dark soliton solution for $\psi_-$-superfluid (BEC-like) at $\frac{T}{T_c}=0.48$.  The behavior of dark solitons and vortices has been known to exhibit many differences in the BEC and BCS regimes of ultracold atomic gases~\cite{GPS,RT,SPRSU}.  Interestingly, these different features also appear in $\psi_\pm$ holographic superfluids~\cite{KKNY0,KKNY,KKNY1,KKNY2}.
}
\label{fig:dsbcs}
\end{figure}

For definiteness we will take the mass square of $\Psi$ to be $m^2=-{2 \over L^2}$. In this case, there are two possible boundary conditions for $\Psi$,  leading to two different types of superfluids which will be denoted respectively as $\psi_\pm$. 
Dark soliton solutions $\psi_S$ in the $\psi_\pm$-superfluid phases can be found by solving equations of motion following 
from~\eqref{lag} with appropriate boundary conditions~\cite{KKNY0,KKNY,KKNY1,KKNY2}. 
They also satisfy~\eqref{p1}--\eqref{p2}. Furthermore, it was observed there that a dark soliton in the $\psi_+$-superfluid ($\psi_-$-superfluid ) has features resembling those of a soliton in the BCS  regime (BEC regime) of ultracold atomic gases. See FIG.~\ref{fig:dsbcs} and its caption.
 While holographic superfluids  cannot be directly mapped to either BEC or BCS regimes, the parallels are nevertheless striking. For ease of comparison, we will below refer to $\psi_\pm$-superfluids  as BCS and BEC-like superfluids, although one should keep the caveats in mind. 

We emphasize that once the parameters and boundary conditions of the gravity description~\eqref{lag} are fixed, the system is fully specified at all scales, including the full set of properties of the superfluid phase such as the critical temperature, temperature dependence of the order parameter, dynamics of dark solitons, and so on.

\smallskip

\noindent {\it Linear instability analysis} 

Now consider small perturbations around a dark soliton solution. 
The calculations are very different for the dissipative GPE and  holographic superfluids, but the conclusions are qualitatively 
similar. We will describe the main results below, leaving calculation details to Appendix~\ref{app:lin}. 

Since the system is translationally invariant along the $y$-direction, 
at the level of linear analysis, we can decompose perturbations in terms of Fourier modes 
$e^{i q y}$ in the $y$-direction. More explicitly, we write 
\be \label{hj}
\psi (t, \vx) = \psi_S (x) + \ep f_{\om, q} (x) e^{-i \om t + i q y}
\ee
with $\ep$ a small parameter. Due to the parity symmetry $y \to - y$,  $q$ and $-q$ have identical behavior, so we will 
restrict to $q \geq 0$. The equations for $f_{\om,q} (x)$ can be reduced to an eigenvalue equation with the
eigenvalue proportional to $\om$. Since $\psi_S$ is odd in $x$, one finds that the even and odd parts of $f_{\om, q} (x)$ decouple. 
The eigenvalues $\om _{e}(q), \om_o (q)$~(respectively for the even and odd parts with appropriate boundary condition on $f_{\om q}$) have a discrete spectrum and are all complex. Note that an $\om$-eigenvalue with a positive imaginary part  leads to exponential time growth in~\eqref{hj} and thus corresponds to an unstable mode. 

At a given temperature $T$, one finds that there exists a $q_c (T)$ such that there is exactly one unstable mode for each $q \in [0, q_c (T))$ and the mode is even in $x$. 
The unstable mode turns out to be pure imaginary $\om_{e}^{(0)} (q) = i \lam_q $ and $\lam_q > 0$, which leads to 
$e^{- i \om^{(0)}_e (q) t} = e^{\lam_q  t}$.  We will denote the eigenfunction $f_{\om, q} (x)$ for the unstable mode as $f_q^{(0)} (x)$. 

 \begin{figure}
\begin{center}
\includegraphics[width=4cm]{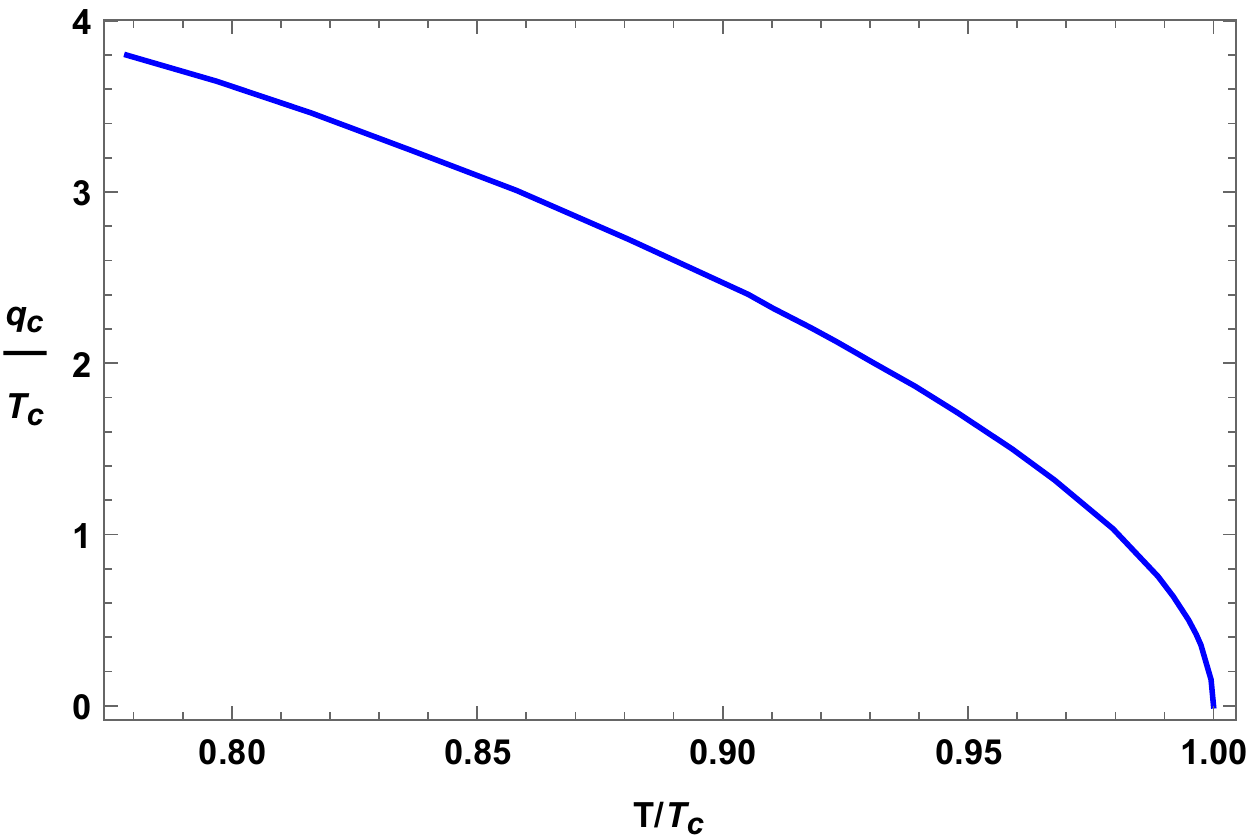}
\includegraphics[width=4cm]{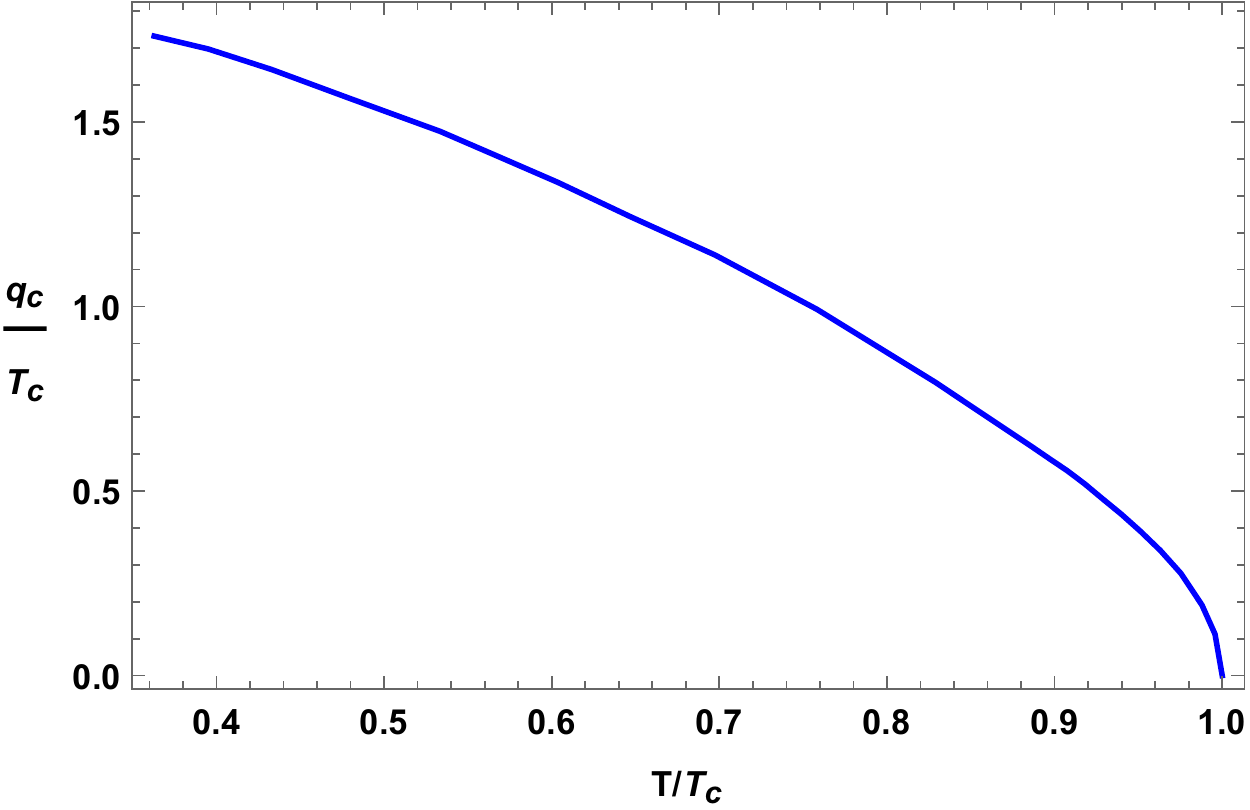}
\end{center}
\caption{The plots of $q_c (T)$  as a function of temperature. The left plot is for $\psi_+$-superfluid (BCS-like), and the right plot is for $\psi_-$-superfluid (BEC-like). 
 }
\label{fig:cutoff}
\end{figure}

The upper value $q_c (T)$ for the unstable mode decreases as one increases the temperature, and for holographic superfluids 
$q_c (T) \to 0$ as $T \to T_c$. More explicitly, we find that as $T \to T_c$ 
\begin{equation}
q_c (T) \approx  \begin{cases}  2.704T_c(1-T/T_c)^{1/2} & \psi_+{ \rm-superfluid} \cr
 0.875T_c(1-T/T_c)^{1/2} &\psi_-{\rm-superfluid} 
 \end{cases} \ .
\end{equation}
See FIG.~\ref{fig:cutoff}. For the dissipative GPE, due to lack of understanding how $\mu$ and $\ka$ depend on $T$, a precise plot is not possible. Nevertheless, assuming that $\mu (T) \approx c (T_c- T)$ as $T_c$ is approached we also find that
$q_c (T) \propto (T-T_c)^\ha$.

We will now show that: (i) the unstable mode at $q=0$ corresponds to self-acceleration; (ii) the unstable mode at $q\neq 0$ corresponds to snake instability; (iii) when a soliton decays via snake instability,  there is a vortex-anti vortex pair to be produced for each
wavelength. For example, consider the system in a finite periodic  box with length $R_y$ along the $y$-direction. The allowed $q$s are thus of the form $q= {2 \pi N \over R_y}$
with $N$ an integer. A snake instability with $q = {2 \pi N \over R_y}$ will then create $N$ pairs of vortices.

To see (i), let us note that  the center $x_c$ of the dark soliton can be identified as being located at the minimum of the condensate, i.e. $\frac{\partial |\psi |^2}{\partial x}|_{x_c}=0$. 
Now consider~\eqref{hj} with $f_{\om, q}$ given by $f^{(0)}_{q=0} (x)$, i.e. $\psi = \psi_S (x) + \ep f^{(0)}_{q=0} (x) e^{\lam_0 t}$, 
which leads to 
\begin{equation}\label{jnl}
|\psi|^2=\psi_S^2+2\epsilon \psi_S (x) {\rm Re}  f^{(0)}_{q=0} (x) e^{\lam_0 t}  \ .
\end{equation}
It can readily seen from the above equation that 
\be 
x_c = - \ep {\eta_0 \ov \xi} e^{\lam_0 t}, {\rm with}~\eta_0 =  {\rm Re}  f^{(0)}_{q=0} (0),~\xi = \p_x \psi_S (0)  \ .
\ee
We thus see that the dark soliton accelerates exponentially.

Now let us consider~\eqref{hj} for a general unstable $q$, with equation~\eqref{jnl} becoming 
\begin{equation}\label{jnj}
|\psi|^2=\psi_S^2+2\epsilon \psi_S (x) {\rm Re} \le(f^{(0)}_{q} (x) e^{i qy} \ri) e^{\lam_q t}  \ .
\end{equation}
We then have 
\be \label{omk}
x_c (t,y) = - \ep {\eta_q \ov \xi} \cos (q y + \th) e^{\lam_q t}, \quad \eta_q = |f^{(0)}_{q} (0)|, 
\ee
and $\th ={\rm arg} f^{(0)}_{q} (0)$. From~\eqref{omk}, $x_c$ depends on $y$ sinusoidally, leading to a snake instability. In particular, for each wave length  ${2 \pi \over q}$ there are two points of zero velocity with opposite circulations around them, corresponding to the locations of a vortex and an anti-vortex to be created. This thus demonstrates (ii) and (iii) above. 
 \begin{figure}
\begin{center}
\includegraphics[width=4cm]{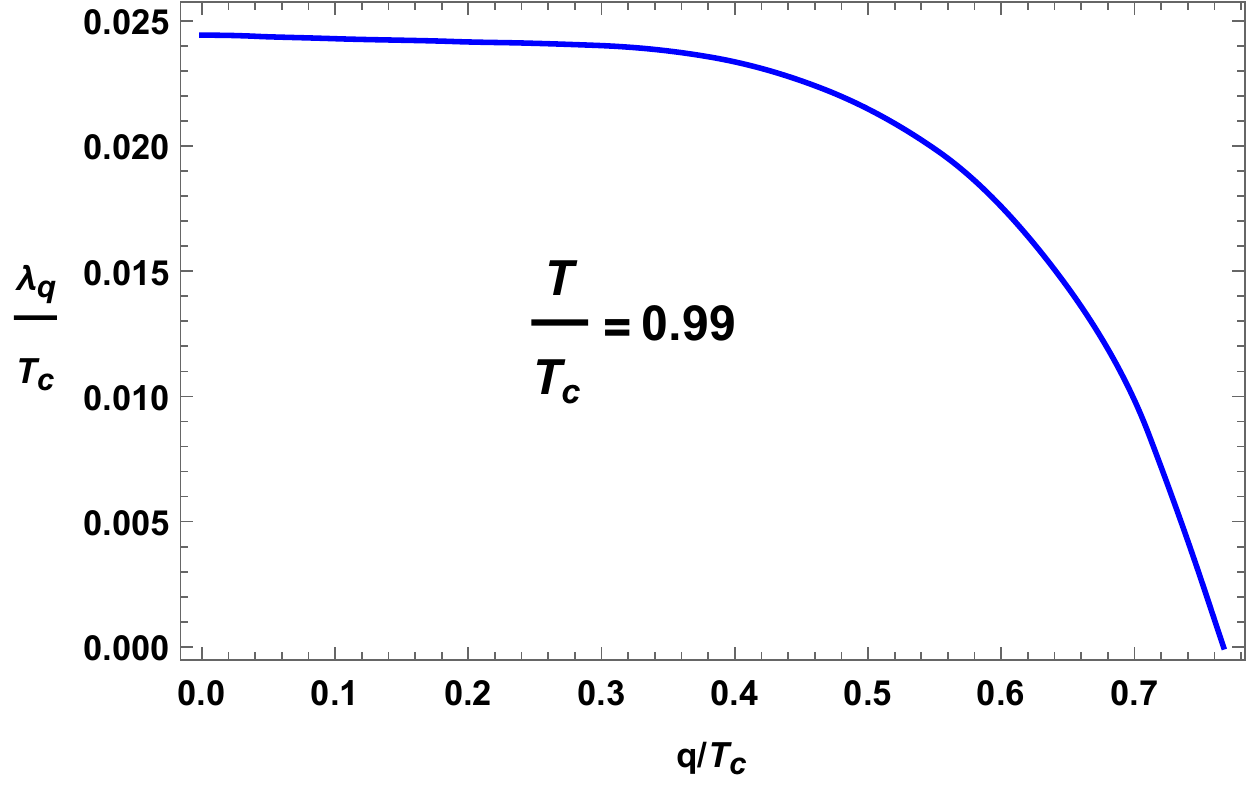}
\includegraphics[width=4cm]{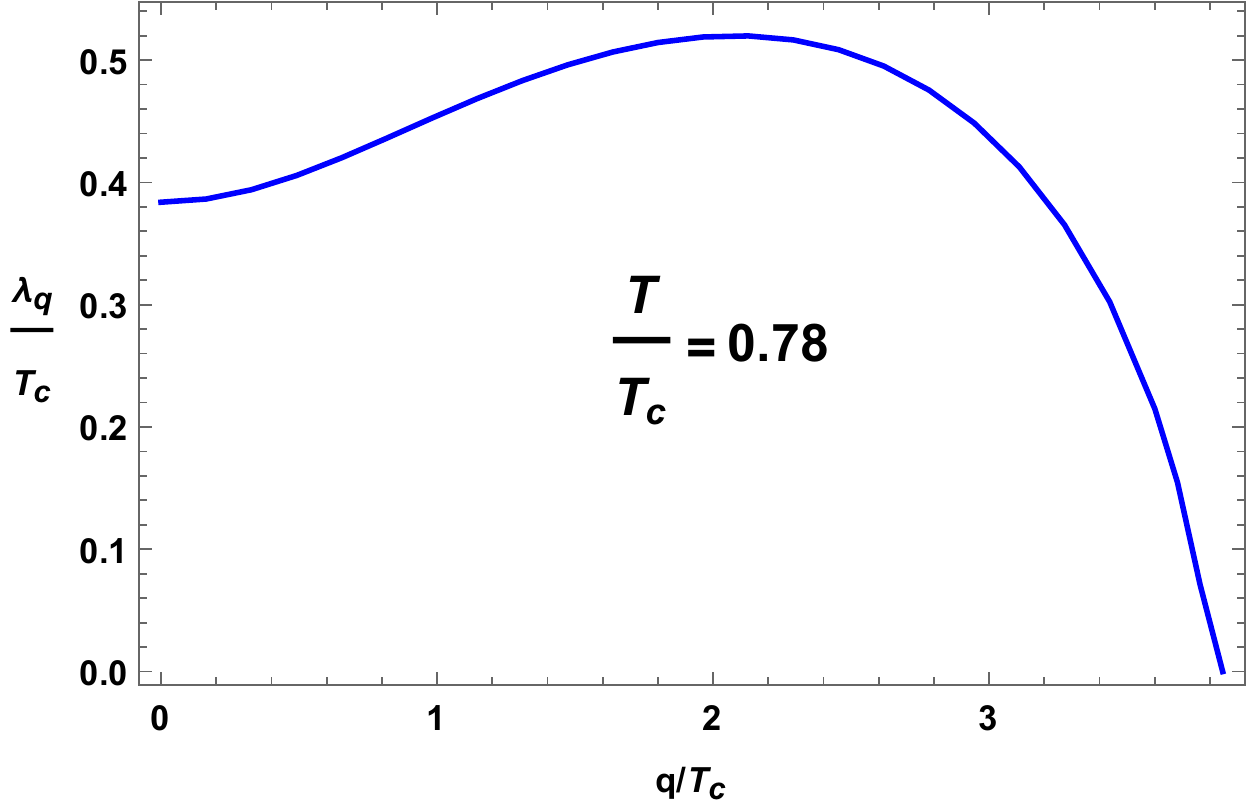}
\includegraphics[width=4cm]{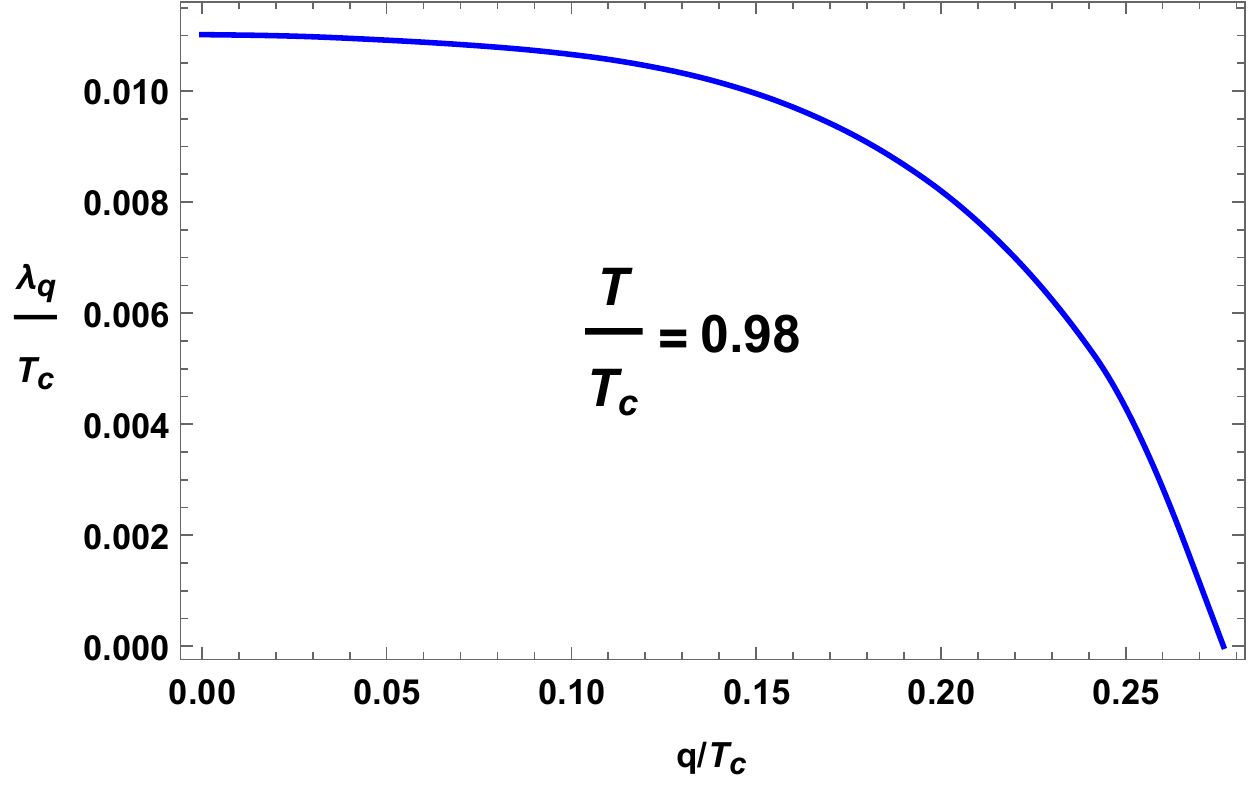}
\includegraphics[width=4cm]{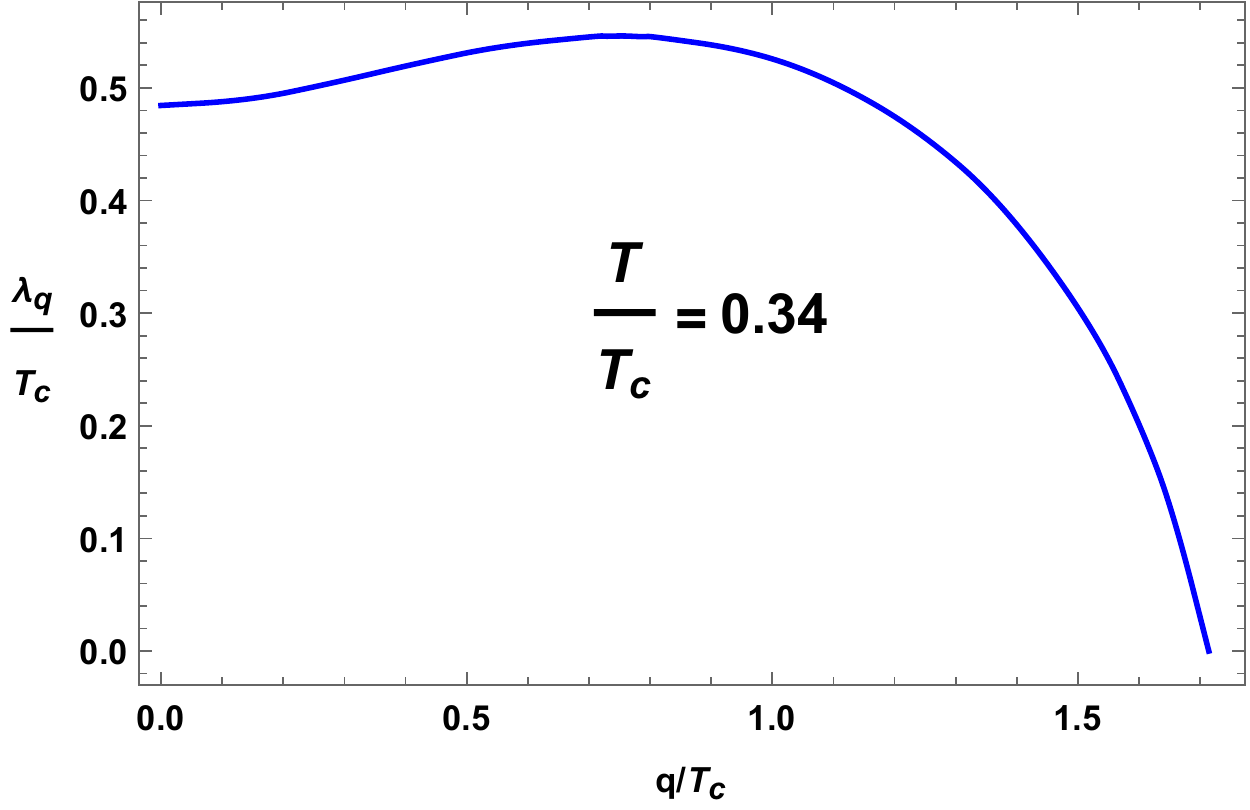}
\end{center}
\caption{The growth exponent $\lam_q$ as a function of $q$. 
The upper panel is for $\psi_+$-superfluid (BCS-like), and the lower panel is for $\psi_-$-superfluid (BEC-like).
Note that in both cases, the maximum lies at $q=0$ at sufficiently high temperature (the left plots), while 
the maximum lies at some $q \neq 0$ at some lower temperature (the right plots). 
}
\label{rate}
\end{figure}

\begin{figure}
\begin{center}
\includegraphics[width=4cm]{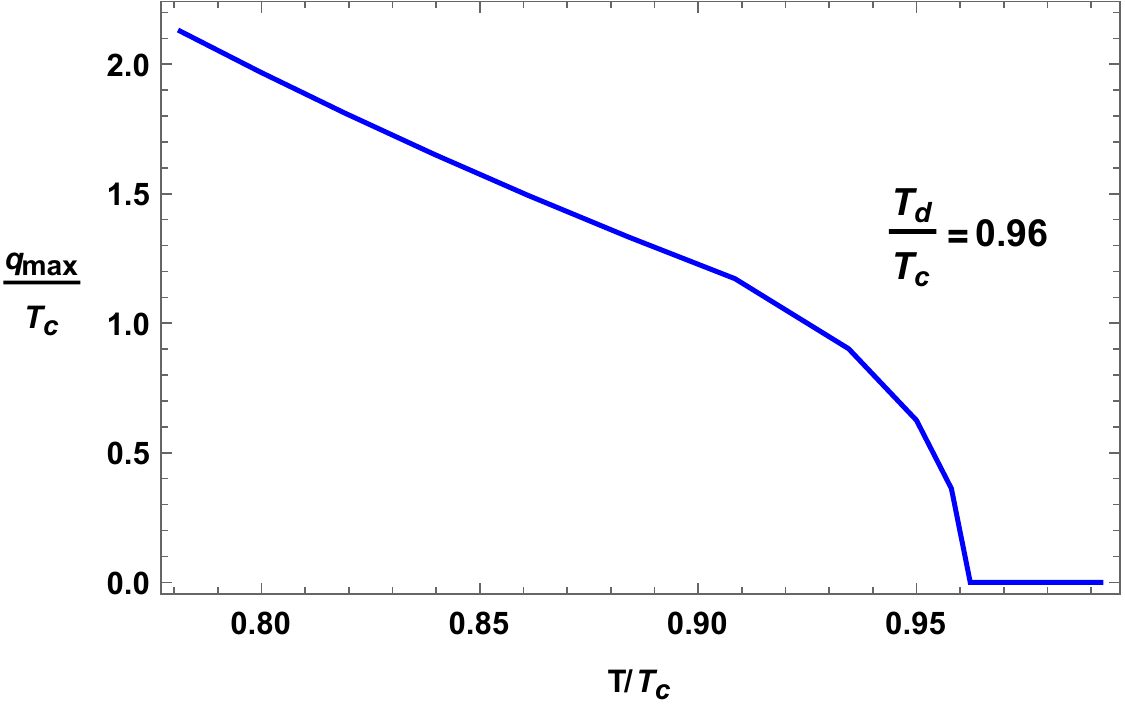}
\includegraphics[width=4cm]{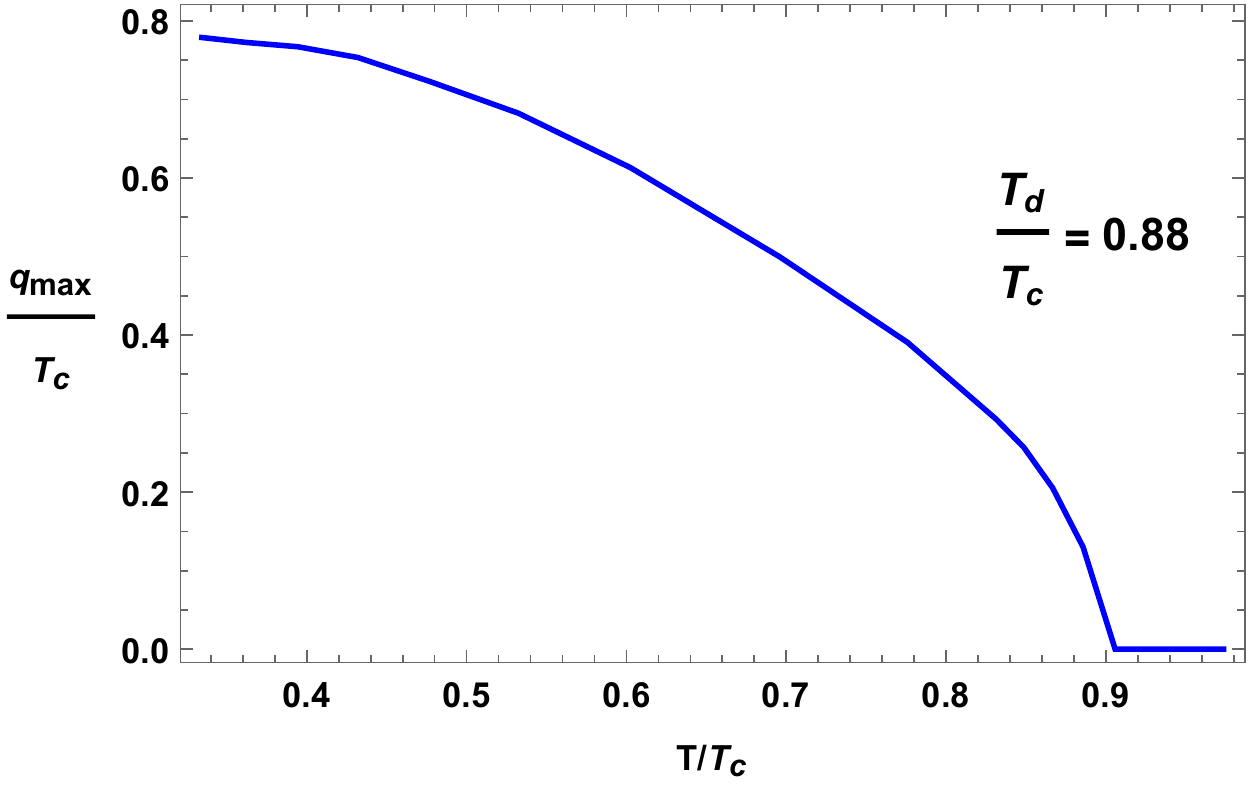}
\end{center}
\caption{$q_{\rm max}$ as a function of temperature. The left plot is for $\psi_+$-superfluid (BCS-like), and the right plot is for $\psi_-$-superfluid (BEC-like). }
\label{fig:dp}
\end{figure}

\medskip

\noindent {\it Switch of dominant decay channel} 

 Among all the unstable modes $q < q_c (T)$, the one with the maximal $\lam_q$ grows fastest and is thus expected 
to be the dominant decay channel of a soliton. 
In FIG.~\ref{rate}, we plot $\lam_q$ as a function $q$ for both holographic superfluids at two different temperatures. 
We notice that the plots for the higher temperature has maximal $\lam_q$ at $q=0$, so the dominant decay channel is the
self-acceleration, while for the lower temperature plots, the maximal value of $\lam_q$ occurs at some nonzero $q$, so the dominant decay channel is the snake instability.  In FIG.~\ref{fig:dp} we plot the value of $q_{\rm max} (T)$ for which maximal value of $\lam_q$ occurs as a function of temperature. We see that $q_{\rm max} (T)$ decreases with temperature monotonically until a value $T_d < T_c$ after which $q_{\rm max}$ becomes identically zero. 

\begin{figure}
\begin{center}
\includegraphics[width=4cm]{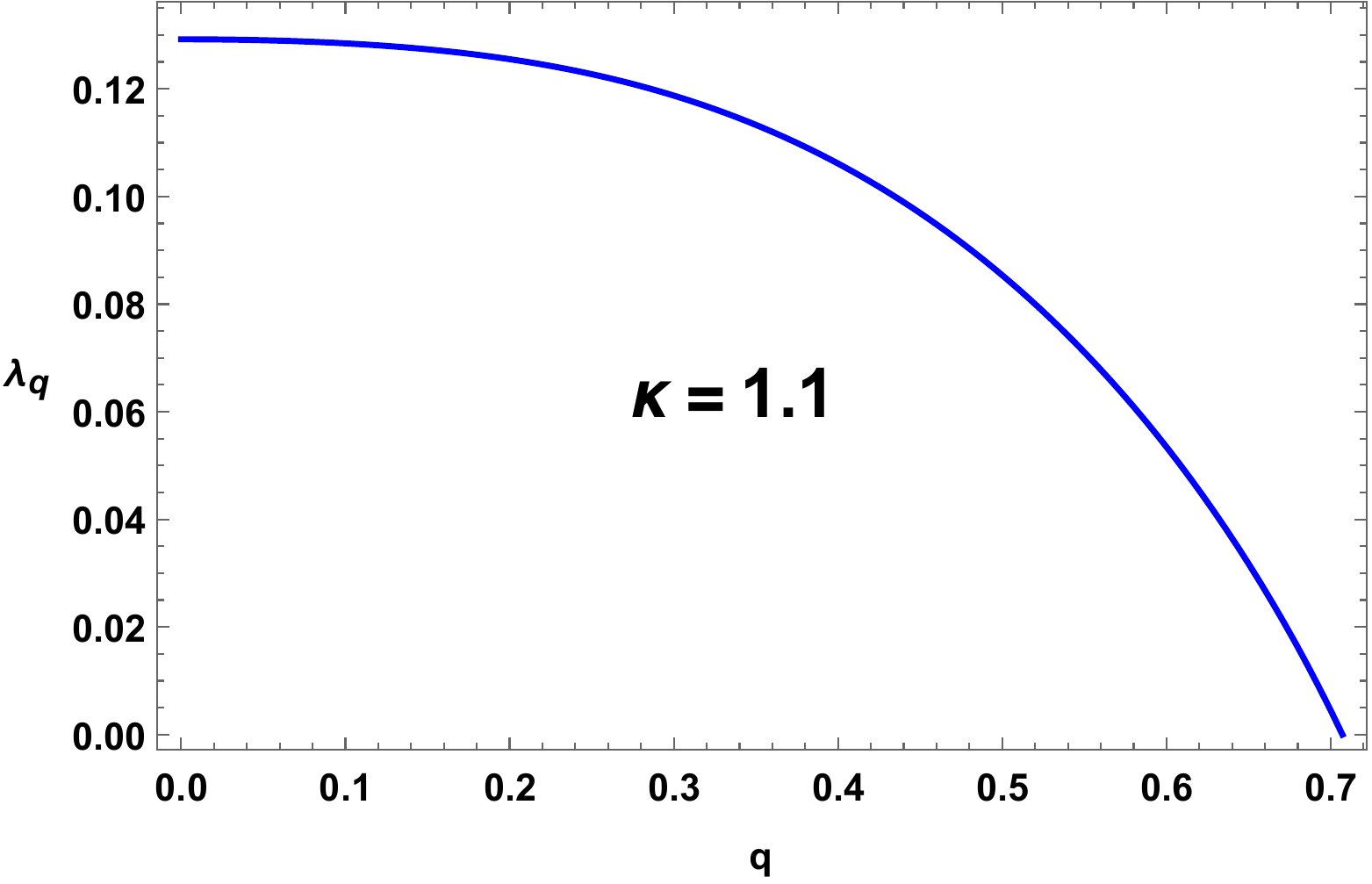}
\includegraphics[width=4cm]{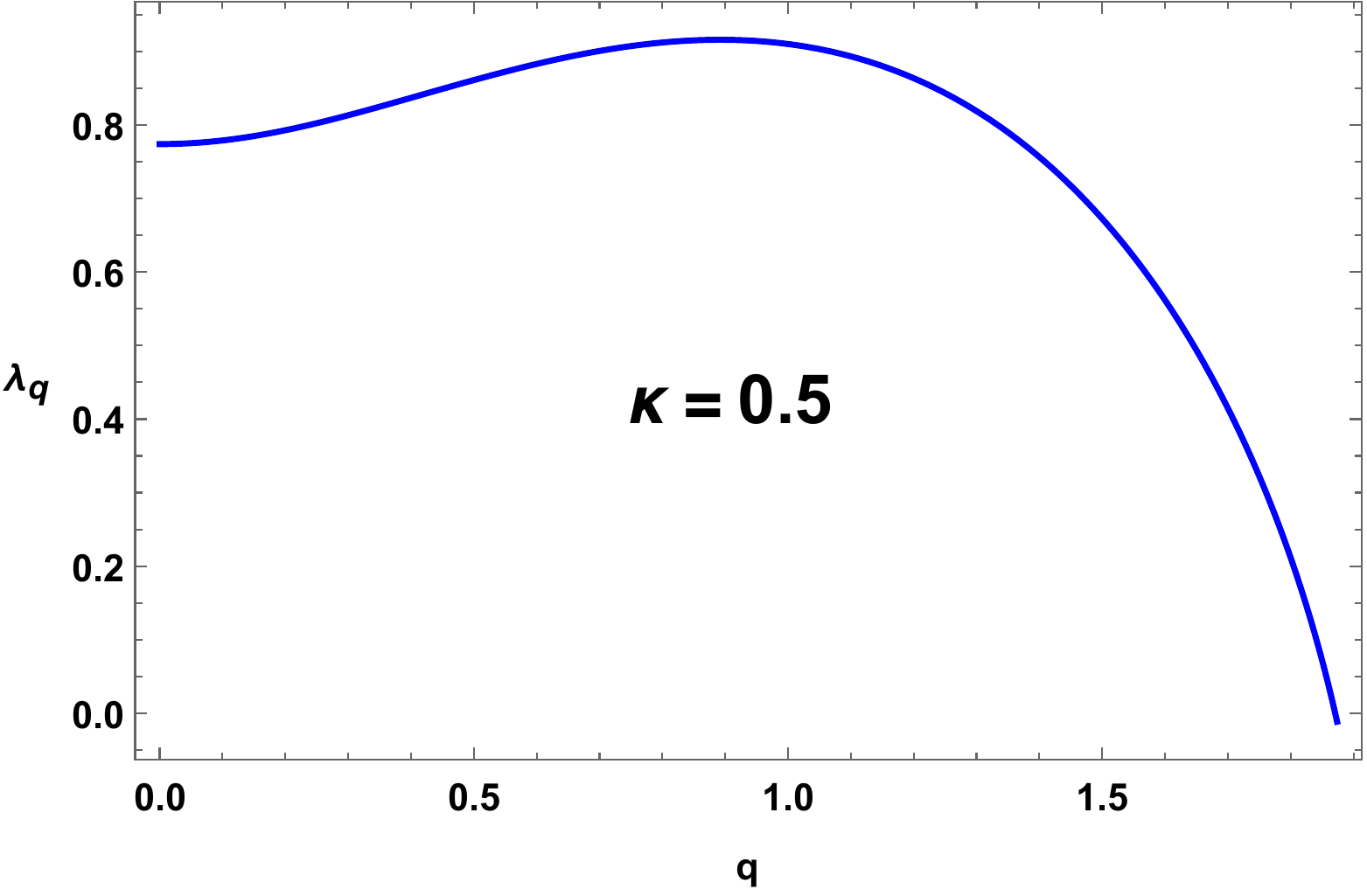}
\end{center}
\caption{Plot of $\lam_q$ as a function of $q$ for the dissipative GPE. The left plot has maximum at $q=0$ while the right plot has maximal at some nonzero $q$.}
\label{fig:gpp}
\end{figure}

\begin{figure}
\begin{center}
\includegraphics[width=6cm]{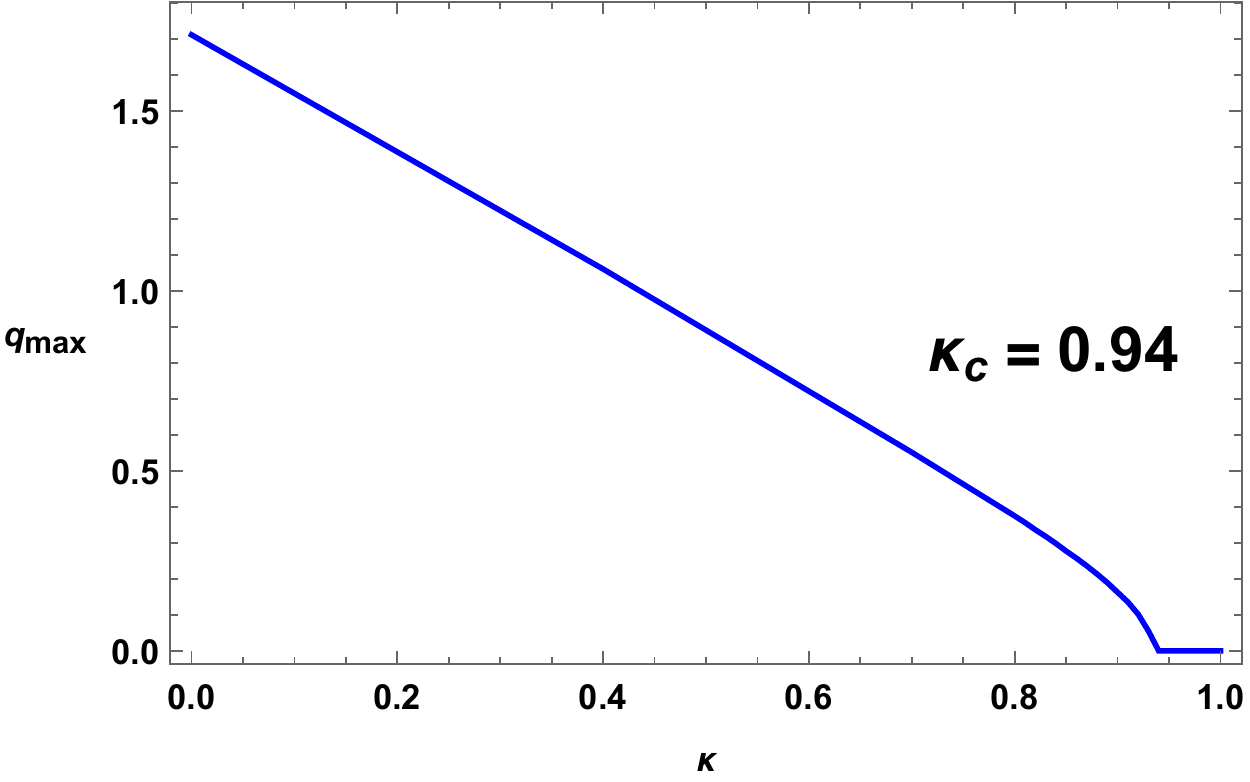}
\end{center}
\caption{Plot of maximal unstable $q_{\rm max}$  as a function of $\kappa$ for the dissipative GPE. In producing the plot we assumed that $\mu(\kappa)=5(1.2-\kappa)$. Similar plot is obtained if one takes $\mu$ to be independent of $\ka$. }
\label{fig:gpe}
\end{figure}

To study this phenomenon for the dissipative GPE, 
we now have to make assumptions on the $\ka$-dependence of $\mu$. We find that the qualitative behavior does not depend sensitively on the choice of $\mu (\ka)$. For example, by choosing $\mu$ to be independent of $\ka$ or to be a linear decreasing function of $\ka$, the very similar behavior is obtained. We present results for the latter in  FIG.~\ref{fig:gpp} and FIG.~\ref{fig:gpe}. 
We thus find that the sharp transition also occurs for the dissipative GPE. 

As $T \to T_d$ from below we find the following ``critical behavior'' 
\be \label{c1}
q_{\rm max} \propto (T_d - T)^\ga  
\ee
with exponent $\ga$ in all cases (holographic superfluids and dissipative GPE)  numerically around $\ha$.
While our current numerical method is not sufficient to make a very accurate determination of $\ga$, there is a simple Ginzburg-Landau type argument which shows that $\ga$ should be given by $\ha$ generically. Consider expanding $\lam_q (T)$ in small $q$ 
\be 
\lam_q (T) = \lam_0 (T) - \ha a (T) q^2 -{1 \ov 4} b (T) q^4 + \cdots 
\ee
Since for $T \approx T_d$, $q_{\rm max}$ is close to zero, and the above expansion should suffice for 
determining the behavior of $q_{\rm max}$ near $T_d$. 
For $T \gtrsim T_d$, $q_{\rm max}$ is zero, we thus should have $a (T) > 0$, while 
for $T \lesssim T_d$, we should generically have $a (T) < 0, b (T_d) > 0$ in order to have $q_{\rm max} \neq 0$. We thus conclude that 
near $T=T_d$ we can expand $a(T)$ as
\be 
a (T) = a_0 (T - T_d) + \cdots , \quad a_0 > 0  \ .
\ee 
It then follows that 
\be 
q_{\rm max} = \sqrt{a_0 (T_d - T) \ov b_0}, \; \; T \lesssim T_d  , \; \;  b_0 = b (T_d) >0 \ .
\ee

From our earlier discussion of self-acceleration and snake instability, we thus conclude that there is a sharp transition at $T_d$ in how dark solitons decay: for $T_c > T > T_d$,  dark solitons  decay by 
self-acceleration, while for $T< T_d$, they decay by snake instability. In particular, in the snake instability regime, the number of vortex and anti-vortex pairs created by the decay of a dark soliton, which is proportional to $q_{\rm max}$, increases as we decrease the temperature. 

\begin{figure}
\begin{center}
\includegraphics[width=4cm]{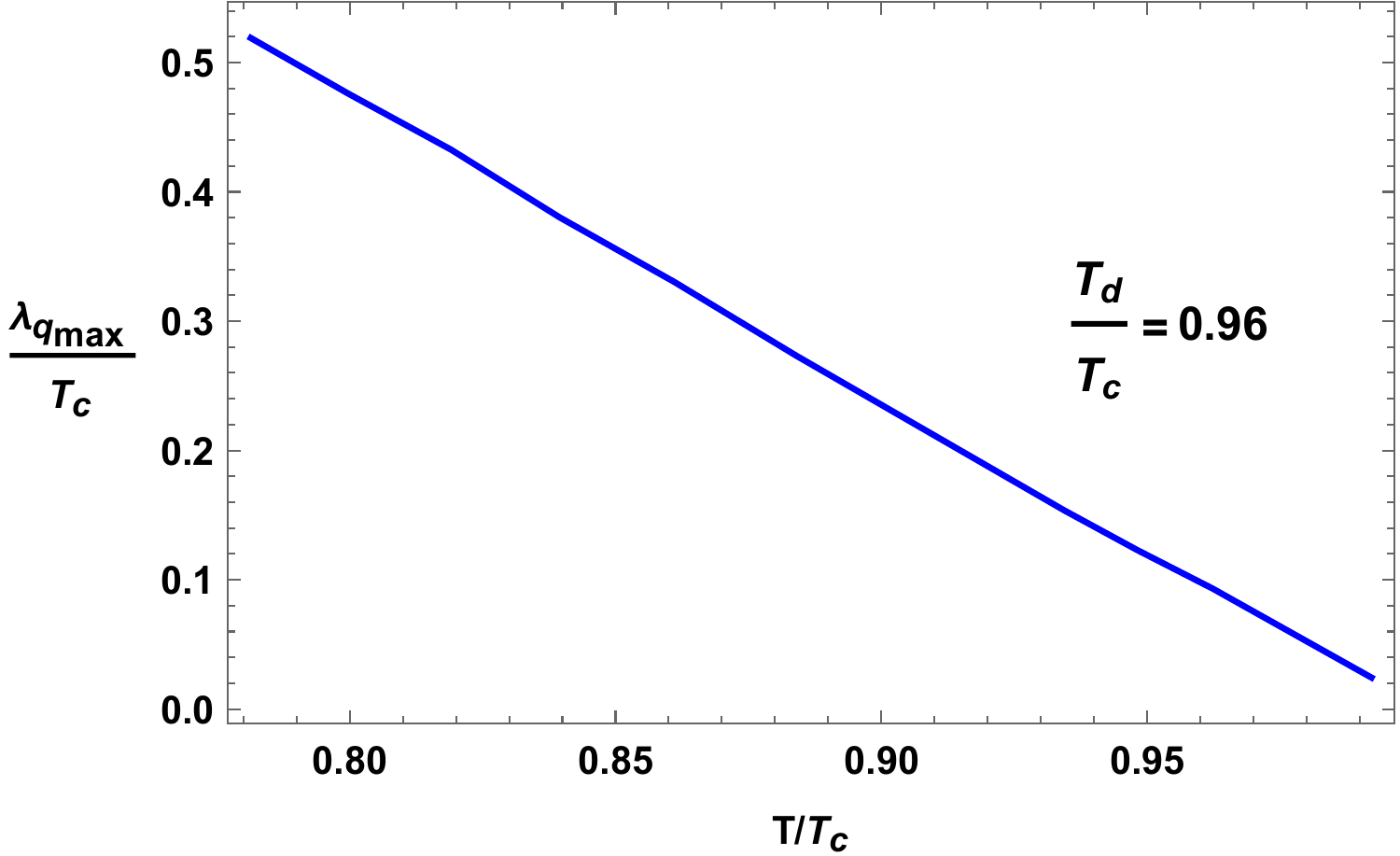}
\includegraphics[width=4cm]{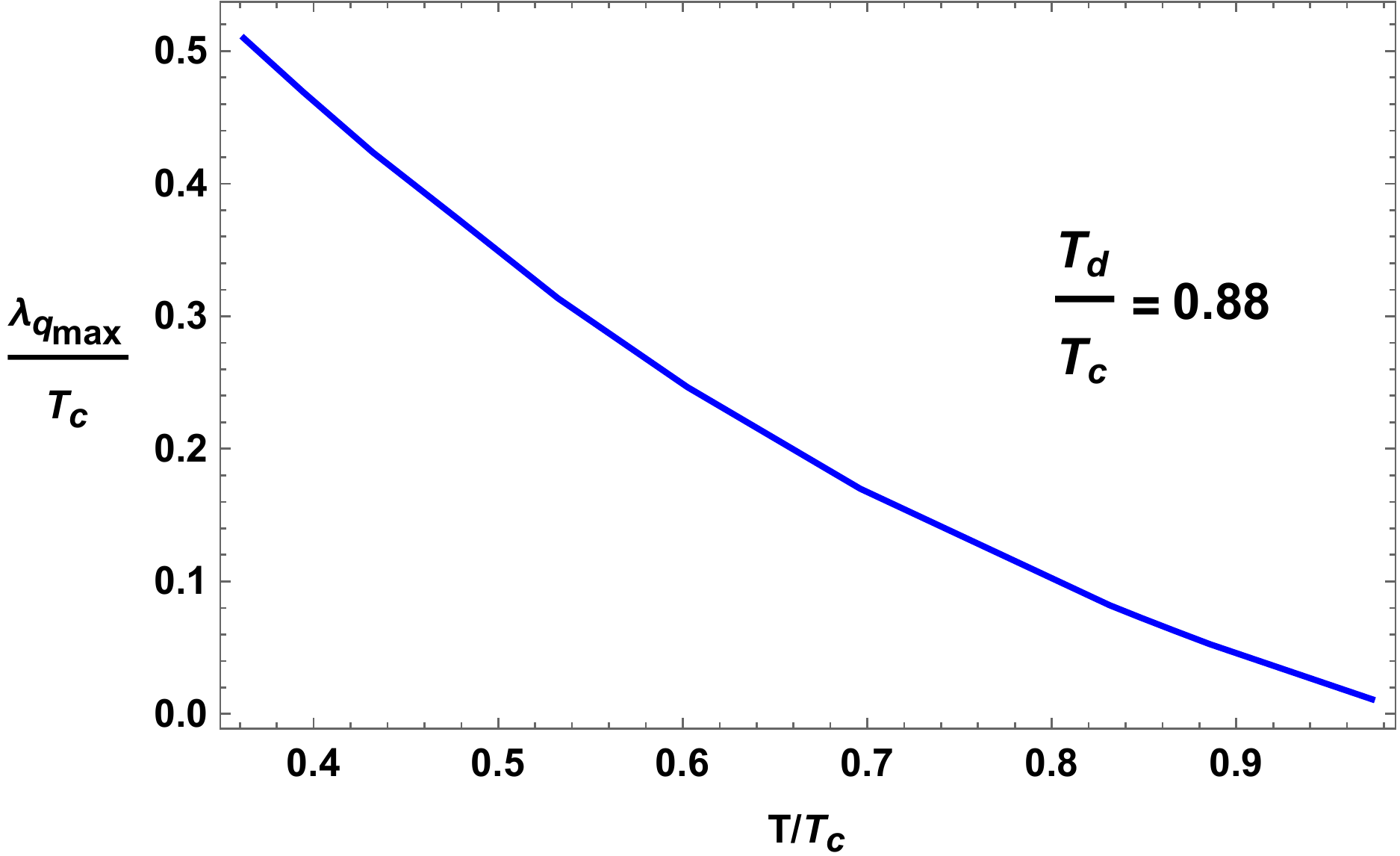}
\includegraphics[width=4cm]{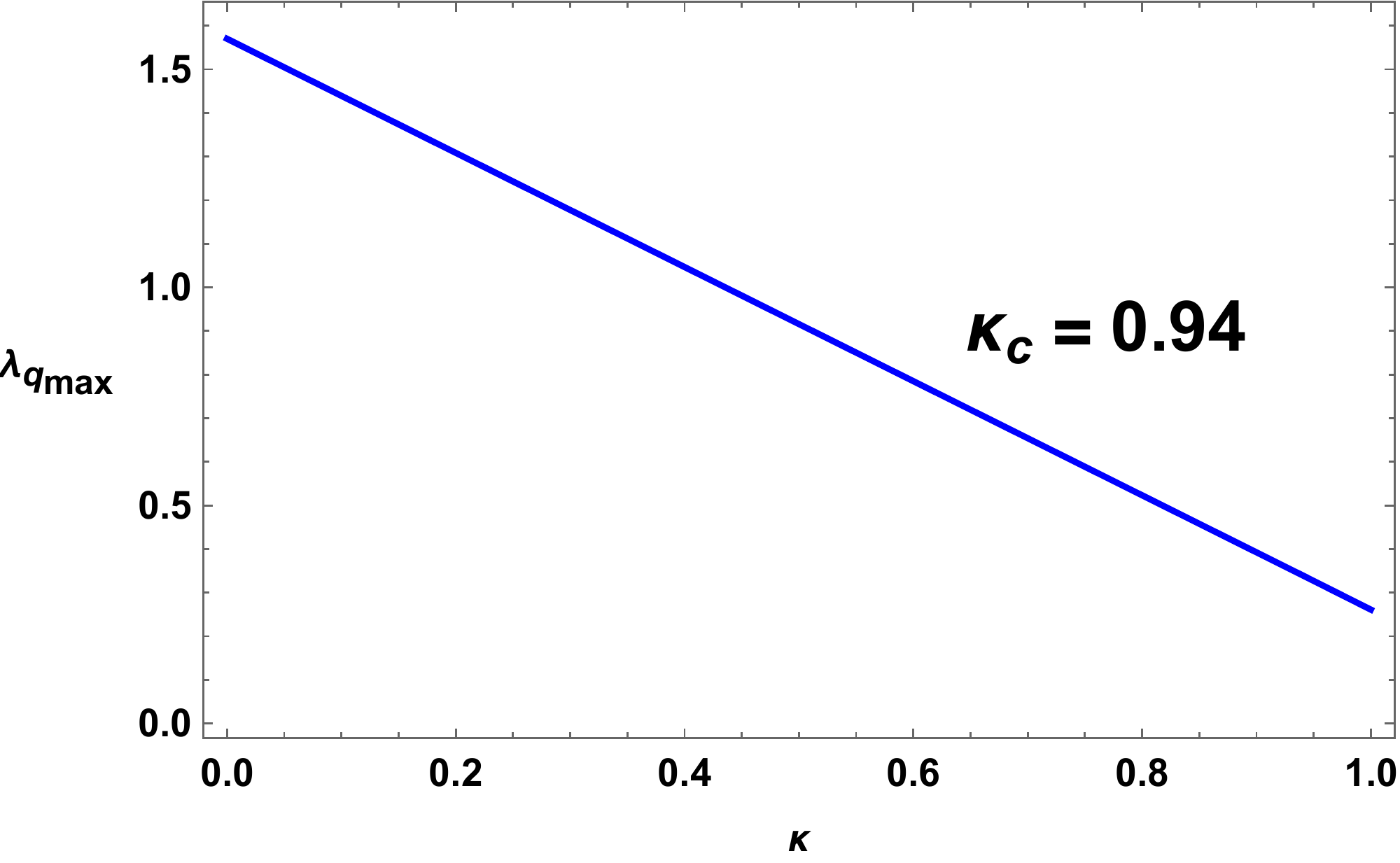}
\end{center}
\caption{$\lam_{q_{\rm max}}$ as a function of temperature. The left plot is for $\psi_+$-superfluid (BCS-like), and the right plot is for $\psi_-$-superfluid (BEC-like), and the bottom plot is for dissipative GPE with $\mu (\ka)$ chosen as in FIG.~\ref{fig:gpe}. 
It is intriguing that for $\psi_+$ and GPE the dependence is almost linear.}
\label{fig:life}
\end{figure}

The value $\lam_{q_{\rm max}}$ characterizes the growth rate of instabilities and thus its inverse may be considered as giving the time scale for the ``lifetime'' of a dark soliton. In FIG.~\ref{fig:life} we plot the temperature dependence of 
 $\lam_{q_{\rm max}}$ for the three types of superfluids. It is interesting to note the higher temperature, the instabilities grow slower 
 and thus the longer lifetime of a soliton.

\medskip

\noindent {\it Dynamical phase transition} 

Now consider a superfluid system in a non-equilibrium state with a finite initial density of dark solitons. 
Then there is a non-equilibrium dynamical phase transition at $T_d$ in how the system relaxes back to equilibrium. 
For $T > T_d$, the solitons directly decay to sound waves which subsequently equilibrate. For $T < T_d$ 
there exists an intermediate phase with an initial density $n_v$ of vortex-antivortex pairs, which is proportional to $q_{\rm max}$.
The lower temperature, the larger $q_{\rm max}$ and $n_v$. Furthermore, from~\eqref{c1} as $T_d$ is approached from below $n_v$ 
has critical behavior
\be 
n_v \propto q_{\rm max} \propto (T_d - T)^\ga , \quad \ga = \ha 
 \ .
\ee

To confirm our expectation that the decay of a dark soliton is controlled by the mode $q_{\rm max}$ with the largest $\lam_q$,
and the dynamical phase transition, we now consider the full  nonlinear evolution of initial states with dark solitons.

\begin{figure}
\begin{center}
 \includegraphics[width=3cm,height=10cm]{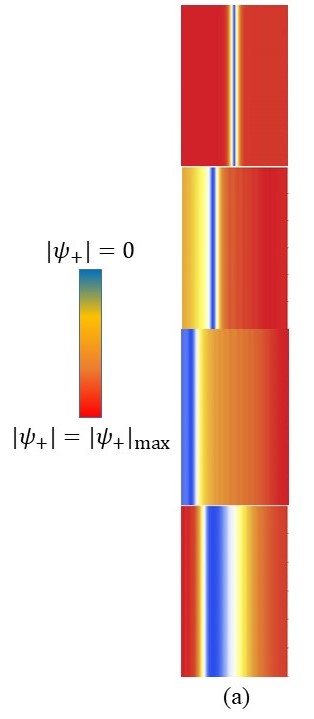}
\includegraphics[width=1.5cm,height=10cm]{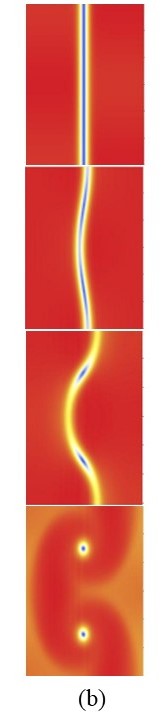}
\includegraphics[width=1.5cm,height=10cm]{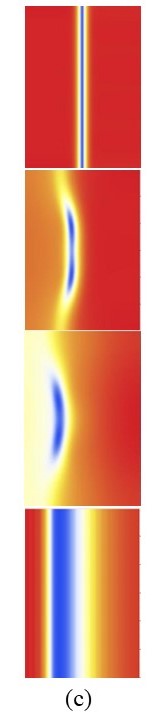}
\includegraphics[width=1.5cm,height=10cm]{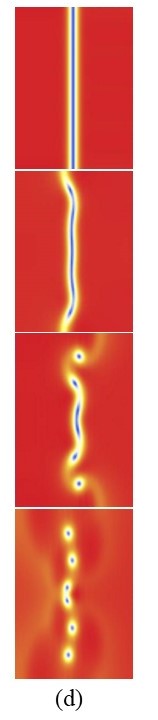}
\end{center}
\caption{The density plots of the time evolution of a dark soliton under various initial perturbations for the $\psi_+$-superfluid (BCS-like). The plots for $\psi_-$-superfluid and dissipative GPE are similar. 
Time increases from top to bottom in each panel.  
The perturbations for (a) and (b) have the form $\delta\psi \propto e^{iqy}$  with (a)  $q=0$ and  (b)  $q=\frac{2\pi}{R_y}$,  at $\frac{T}{T_c}=0.78$. The perturbations for (c) and (d) are of the form $\de \psi \propto \sum_q e^{i \al_q} e^{i q y}$ with random phases $\al_q$ with (c) at $\frac{T}{T_c}=0.99$ and (d) at $\frac{T}{T_c}=0.78$.}
\label{pure}
\end{figure}

Let us first consider the case of a single initial dark soliton under the following two types of perturbations: (i)
a single-wave length perturbation $\delta\psi \propto e^{iqy}$ 
with $q \in [0, q_c (T))$, and (ii)  a general perturbation of the form 
$\de \psi \propto \sum_q e^{i \al_q} e^{i q y}$ with random phases $\al_q$. For (i), with $q=0$,  one finds that different parts of the dark soliton accelerate uniformly. During its acceleration, the dark soliton also broadens, and eventually dissolves into sound waves. For $q\neq 0$, the acceleration pattern for different parts of the soliton shows sinusoidal behavior with wave length ${2 \pi \ov q}$, consistent with  the prediction of~\eqref{omk}. In particular, the vortex-antivortex formation as well as their numbers are 
precisely as predicted below~\eqref{omk}. See FIG.~\ref{pure} (a) (b).
For a general perturbation (ii),  the resulting evolutions are shown in FIG.~\ref{pure} (c) (d). We indeed find that at a temperature above $T_d$, the soliton accelerates without any vortex formation~(FIG.~\ref{pure} (c)). At a temperature below $T_d$, the evolution is dominated by snake instability, evidenced by formation of vortex pairs.  Furthermore, the number of vortex pairs is consistent with linear analysis. For instance, for $\psi_+$-superfluid at $\frac{T}{T_c}=0.78$, the linear analysis tells us that the mode $q=\frac{2\pi N}{R_y}$ with $N=3$ is the most unstable mode. We find the non-linear evolution indeed produces $3$ vortex pairs (see FIG.~\ref{pure} (d)). 

\begin{figure}
\begin{center}
\includegraphics[width=11cm,height=10cm]{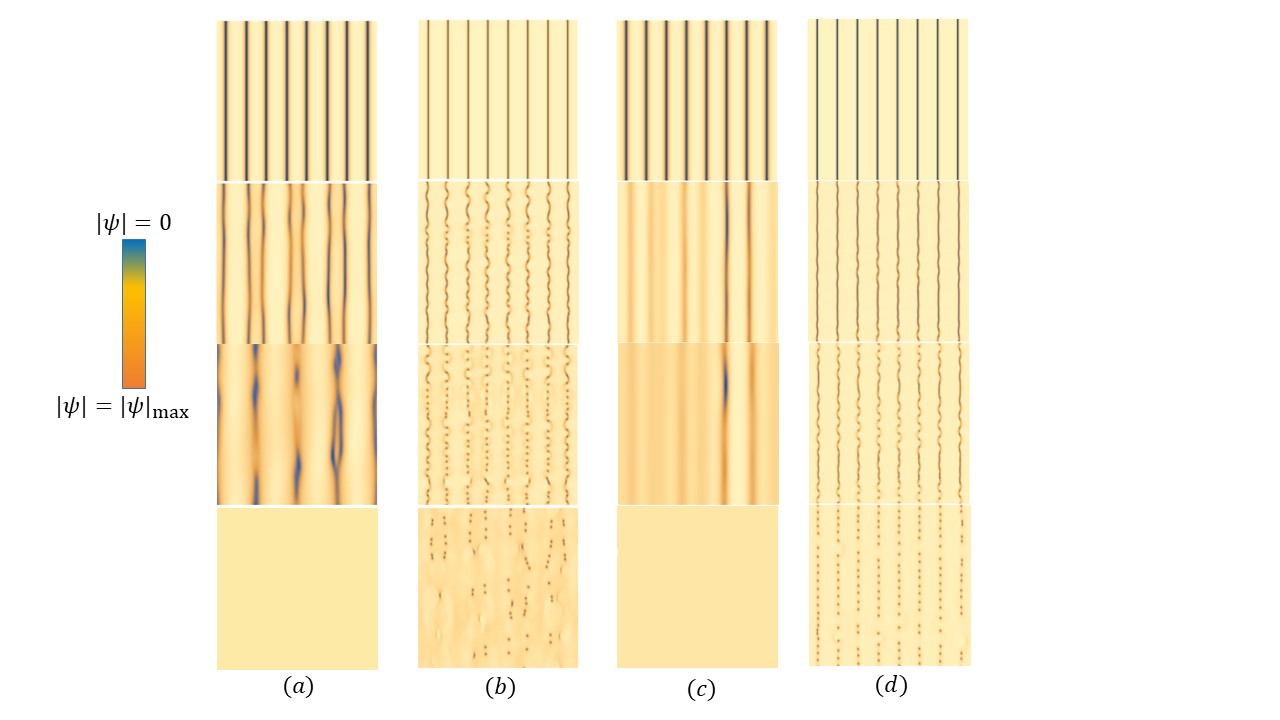}
\end{center}
\caption{The density plots of the time evolution of a finite density of dark solitons under generic initial perturbations. 
For simplicity we have taken the initial solitons to be parallel to one another with equal distances rather than randomly distributed as would be in a generic non-equilibrium initial state. Panels (a) (b) are for $\psi_+$-superfluid (with $T_d = 0.96 T_c$) at temperatures $\frac{T}{T_c}=0.99$ and $\frac{T}{T_c}=0.78$ respectively, while panels (c) (d) are for dissipative GPE (with $\ka_d = 0.94$) at $\ka =1.1$ and $\ka = 0.4 $ respectively. Time increases from top to bottom in each panel. In the bottom plot for panel (b), the vortex pairs have already partially annihilated after creation. 
}
\label{fig:multi}
\end{figure}

Finally we consider the full nonlinear evolution of a system with a finite initial density of dark solitons. On physical ground, we expect our results regarding the stability of a single soliton should apply to a finite density of solitons as far as the 
density is not too high, i.e. as far as the average distances between solitons are larger than the healing length of a soliton (which is often treated as a microscopic scale). This expectation is indeed confirmed by full nonlinear simulations, see FIG.~\ref{fig:multi} for contrast of two temperatures below and above $T_d$.  Thus nonlinear evolutions fully confirm the existence of a dynamical phase transition. 


\medskip

\noindent {\it Discussion}

To summarize, by studying the decay of dark solitons, we identified a novel dynamical phase transition
which results from temperature dependence of non-equilibrium dynamics of solitons. 
The phase transition involves the appearance of an intermediate (unstable) phase 
with the order parameter given by the density of vortex pairs and a critical exponent $\ga = \ha$.
This vortex gas/liquid phase could have important implications for understanding non-equilibrium dynamics of a superfluid, 
for example, it could be long-lived enough to exhibit vortex and wave turbulent states. 

It would be extremely interesting to search for such a dynamical phase transition in experiments, and to measure the critical temperature $T_d$ as well as the critical exponent $\ga$. The fact that we observe the phase transition in such vastly different theories as the dissipative GPE and holographic systems gives much confidence in that it is a universal phenomenon. A natural place to look for such a phase transition is ultracold atomic gases which we expect should exhibit this phenomenon both in the BEC and BCS regimes. The transition should be already detectable  in the usual experimental setup of a conventional harmonic potential. Recent experimental advances in the realization of box-like optical traps~\cite{GSGSH} can make the comparisons
with our theoretical results even more direct. 

At a technical level, our approach to decay of dark solitons has a few 
important new features compared with earlier studies~\cite{MLS,BA,BA1,GK,CBSDP,CMB,FMS,MSESL,PPBA,JPB,GNZ,Cockburn,MR,LAKT},  Our linear instability analyses not only provide a unified treatment for self-acceleration\footnote{Note that the argument in~\cite{BA,BA1} for self-acceleration used perturbations which do not vanish as $x \to \pm \infty$. This is unphysical, as that requires changing the asymptotic behavior of the system. In contrast, our unstable eigenmodes are all local, going to zero exponentially fast as $x \to \pm \infty$.} 
and snake instabilities, but also give a wealth of information on temperature dependence of dominant decay channels, the number of vortex pairs produced, as well as time scale for the lifetime of a soliton.
Our full nonlinear evolution of the decay of a dark soliton at finite temperatures provide strong support for the conclusions of linear analyses. Nonlinear evolution of a dark soliton was studied previously in~\cite{CBSDP} using the time-dependent BdG equations of the BCS-BEC crossover at zero temperature, where snake instability was 
observed, but not self-acceleration. This is consistent with our general conclusion that at low temperatures snake instabilities should dominate. 




{\it Acknowledgements.}--M.G. is partially supported by NSFC with Grant No.11675015, 11775022, and 11875095, as well as by CSC. He also thanks the Perimeter Institute “Visiting Graduate Fellows” program. Research at Perimeter Institute is supported by the Government of Canada through the Department of Innovation, Science and Economic Development and by the Province of Ontario through the Ministry of Research, Innovation and Science. E.K-V. is partially supported by the Academy of Finland grant no 1297472, and by a grant from the Vilho, Yrj\"o and Kalle V\"ais\"al\"a\ Foundation.
 H.L is partially supported by the Office of High Energy Physics of U.S. Department of Energy under grant Contract Number DE-SC0012567. Y.T. is partially supported by NSFC with Grant No.11975235 and he is also supported by the Strategic Priority Research Program of the Chinese Academy of Sciences with Grant No.XDB23030000.  H.Z. is supported in part by FWO-Vlaanderen through the project G006918N, and by the Vrije Universiteit Brussel through the Strategic Research Program ``High-Energy Physics". He is also an individual FWO fellow supported by 12G3515N.


\newpage

\appendix


\section{Numerical scheme for a holographic dark soliton construction and its fully non-linear simulation by perturbations }  \label{app:b}

For simplicity but without loss of generality, we will focus only on the case of $m^2L^2=-2$ in the axial gauge $A_z=0$, in which the equations of motion on top of the background~\eqref{ui} can be written in an explicit way as\cite{LTZ}
\begin{eqnarray}\label{evolution1}
 \partial_t\partial_z\Phi &=& iA_t\partial_z\Phi+\frac{1}{2}[i\partial_zA_t\Phi+f\partial_z^2\Phi+f^\prime\partial_z\Phi\nonumber\\
  &&+(\mathbf{\partial}-i\mathbf{A})^{2}\Phi-z\Phi],\\
  \label{evolution2} \partial_t\partial_z\mathbf{A} &=& \frac{1}{2}[\partial_z(\mathbf{\partial}A_t+f\partial_z\mathbf{A})+(\mathbf{\partial}^2\mathbf{A}-\mathbf{\partial}\mathbf{\partial}\cdot\mathbf{A})\nonumber\\
  &&-i(\bar{\Phi}\mathbf{\partial}\Phi-\Phi\mathbf{\partial}\bar{\Phi})]-\mathbf{A}\bar{\Phi}\Phi, \\
\label{constraint} \partial_z^2A_t&=&\partial_z\mathbf{\partial}\cdot\mathbf{A}+i(\bar{\Phi}\partial_z\Phi-\Phi\partial_z\bar{\Phi}),\\
\label{evolution3}    \partial_t\partial_zA_t &=& \mathbf{\partial}^2A_t+f\partial_{z}\mathbf{\partial}\cdot\mathbf{A}-\partial_t\mathbf{\partial}\cdot\mathbf{A}-2A_t\bar{\Phi}\Phi \nonumber\\
 && +i f(\bar{\Phi}\partial_{z}\Phi-\Phi\partial_{z}\bar{\Phi})-i(\bar{\Phi}\partial_t\Phi-\Phi\partial_t\bar{\Phi})
 \end{eqnarray}
 with $\Phi=\frac{\Psi}{z}$\cite{units}.
The asymptotic solution of $A$ and $\Phi$ can be expanded near the AdS boundary as
\begin{equation}\label{near}
A_\nu=a_\nu+b_\nu z+o(z),\quad\Phi=\phi_1+\phi_2 z+o(z).
\end{equation}
Then the expectation value of $J$ and $\psi_\pm$ (the condensate of $\psi_\pm$-superfluids) can be explicitly obtained by holography as the variation of the renormalized bulk on-shell action with respect to the sources, {\em i.e.},
\begin{eqnarray}
\langle J^\nu\rangle&=&\frac{\delta S_\pm}{\delta a_\nu}=\lim_{z\rightarrow 0}\sqrt{-g}F^{z\nu},\label{current}\\
\langle \psi_+\rangle&=&\frac{\delta S_+}{\delta \phi_1}=\bar{\phi}_2-\dot{\bar{\phi}}_1-ia_t\bar{\phi}_1,\label{vev1}\\
\langle \psi_-\rangle&=&\frac{\delta S_-}{\delta \langle \psi_+\rangle}=-\bar{\phi}_1,\label{vev2}
\end{eqnarray}
where the dot denotes the time derivative, and $S_\pm$ is the renormalized action, obtained by adding the corresponding counter term to the original action to make it finite and well posed for the variational principle as \cite{LTZ1}
\begin{eqnarray}
S_+&=&S-\int_\mathcal{B}\sqrt{-\gamma}|\Psi|^2,\\
S_-&=&S+\int_\mathcal{B}\sqrt{-\gamma}[|\Psi|^2+(n^a\overline{D_a\Psi}\Psi+c.c.)].
\end{eqnarray}

In~\cite{KKNY0,KKNY}, the dark soliton was found in  the Schwarzschild coordinates by the finite difference scheme with the relaxation method. Here we will instead reconstruct such a dark soliton solution in  Eddington coordinates~\eqref{ui} by the pseudo-spectral scheme with the Newton-Raphson method. This turns out to be much more efficient, and enables us to study  lower temperatures. To achieve this, we make the following ansatz for the non-vanishing bulk fields,
\begin{equation}
A_t=A_t(z,x), \Phi=\varphi(z,x) e^{i\theta(z,x)}
\end{equation}
with
\begin{equation}
\partial_z\theta=-\frac{A_t}{f}, A_x=\partial_x\theta.
\end{equation}
Then the equations of motion reduce to
\begin{eqnarray}
0&=&f\partial_z^2\varphi+f'\partial_z\varphi+\partial_x^2\varphi+\frac{A_t^2}{f}\varphi-z\varphi,\\
0&=&f\partial_z^2A_t+\partial_x^2A_t-2\varphi^2A_t.
\end{eqnarray}
Associated with the superfluid condensate $\langle \psi_\pm\rangle$, the dark soliton solution can thus be obtained by imposing
the boundary conditions
\begin{equation}
A_t=\mu, \varphi=0~(\partial_z\varphi=0), \theta=0
\end{equation}
at the AdS boundary, as well as the Neumann boundary conditions at $x=\pm\frac{R_x}{2}$, where $R_x$ is set to be much larger than the characteristic length of the dark soliton in consideration, so that the finite size effect can be removed. 

On the other hand, for our purpose, the fully non-linear simulation starts with the initial data 
\begin{equation}
\Phi=\Phi_S+\epsilon\delta\Phi, \mathbf{A}=\mathbf{A}_S+\epsilon\delta\mathbf{A},
\end{equation}
where the subscript $S$ denotes the corresponding profile for the static dark soliton solution. For simplicity but without loss of generality, we take $\delta\Phi=ze^{iqy}$ or a random superposition of modes with different $q$s, and $\delta\mathbf{A}=0$~\cite{periodic}. Then $A_t$ can be solved by the constraint equation (\ref{constraint}) subject to the  boundary conditions
\begin{equation}\label{bc}
A_t=\mu, \partial_zA_t=-n_S
\end{equation}
at the AdS boundary. Next, we can evolve $\Phi$ and $\mathbf{A}$ through Eq.(\ref{evolution1}) and Eq.(\ref{evolution2}) subject to the source free boundary condition at the AdS boundary, and the boundary conditions 
\begin{equation}
\partial_x\Phi=0, \mathbf{A}=0
\end{equation}
at $x=\pm\frac{R_x}{2}$. Furthermore, we can evolve the second boundary condition in Eq.(\ref{bc}) by evaluating Eq.(\ref{evolution3}) at the AdS boundary, i.e.,
\begin{equation}
\partial_t\partial_zA_t=\partial_z\mathbf{\partial}\cdot\mathbf{A},
\end{equation}
which is essentially the conservation law of the boundary particle current. The later time behavior for $A_t$ can be obtained in the same way as described before.

The above evolution scheme is implemented numerically by the fourth order Runge-Kutta method along the time direction, together with the pseudo-spectral method along the space directions, where Chebyshev Polynomials are used in the $z$ and $x$ directions, while Fourier modes are used in the $y$ direction.
This gives a computational advantage over the previous holographic investigations, which deal exclusively with periodic boundary conditions along both the $x$ and $y$ directions \cite{LTZ,ACL,EGKS,DNTZ,CML} and therefore cannot properly accomodate the dynamics of dark solitons.

\section{Linear instability around a dark soliton} \label{app:lin}

Now let us turn to the linear perturbation analysis of our holographic superfluid on top of the dark soliton background. To this end, we would first like to decompose $\Phi$ into its real and imaginary parts as
\begin{equation}
\Phi=a+ib.
\end{equation}
Then taking into account the translation invariance of our dark soliton background along both the time direction and the $y$ direction, we can take the bulk perturbation fields as the form of 
 $\epsilon\delta(z,x)e^{-i\omega t+iqy}$. The resulting linear perturbation equations can be written explicitly as
\begin{eqnarray}
0&=&(q^2+z+A_x^2)\delta a+2(aA_x-\partial_x b)\delta A_x-2A_x\partial_x\delta b\nonumber\\
&&-\partial_x^2\delta a
+(\partial_z A_t-\partial_x A_x)\delta b+2\partial_z b\delta A_t\nonumber\\
&&+(3z^2-2i\omega)\partial_z\delta a-i q b\delta A_y\nonumber\\
&&-b\partial_x\delta A_x+b\partial_z\delta A_t+2A_t\partial_z\delta b+(z^3-1)\partial_z^2\delta a,\\
0&=&(q^2+z+A_x^2)\delta b+2(\partial_x a+A_x b)\delta A_x+2A_x\partial_x\delta a\nonumber\\
&&-\partial_x^2\delta b
+(\partial_x A_x-\partial_z A_t)\delta a-2\partial_z a\delta A_t\nonumber\\
&&+(3z^2-2i\omega)\partial_z\delta b+i q a \delta A_y\nonumber\\
&&+a\partial_x\delta A_x-a\partial_z\delta A_t-2A_t\partial_z\delta a+(z^3-1)\partial_z^2\delta b,\\
0&=&(2a^2+2b^2+q^2)\delta A_x+2(2b A_x+\partial_x a)\delta b\nonumber\\
&&-2\partial_x b\delta a
+2b\partial_x\delta a
+i q\partial_x\delta A_y+4a A_x\delta a\nonumber\\
&&-2a\partial_x\delta b
+(3z^2-2i\omega)\partial_z\delta A_x\nonumber\\
&&-\partial_z\partial_x\delta A_t+(z^3-1)\partial_z^2\delta A_x,\\
0&=&2i q b\delta a+2(a^2+b^2)\delta A_y-2iqa\delta b\nonumber\\
&&-\partial_x^2\delta A_y+iq\partial_x\delta A_x
-iq\partial_z\delta A_t\nonumber\\
&&+(3z^2-2i\omega)\partial_z\delta A_y+(z^3-1)\delta_z^2\delta A_y,\\
0&=&2\partial_z a\delta b-2\partial_z b\delta a+2b\partial_z\delta a+i q\partial_z \delta A_y-2a\partial_z\delta b\nonumber\\
&&+\partial_z\partial_x\delta A_x-\partial_z^2\delta A_t\\
0&=&(q^2+2a^2+2b^2)\delta A_t+q\omega\delta A_y-i\omega\partial_x\delta A_x-\partial_x^2\delta A_t\nonumber\\
&&+2(z^3-1)\partial_z a\delta b-2(z^3-1)\partial_z b\delta a-2i\omega b\delta a\nonumber\\
&&+4A_t b\delta b
+2b(z^3-1)\partial_z\delta a\nonumber\\
&&-i\omega\partial_z\delta A_t+i q(z^3-1)\partial_z\delta A_y+4A_t a\delta a\nonumber\\
&&+2i\omega a\delta b-2a(z^3-1)\partial_z\delta b+(z^3-1)\partial_z\partial_x\delta A_x.\nonumber\\
\end{eqnarray}
Note that the gauge transformation
\begin{equation}
A\rightarrow A+\nabla\vartheta, \Psi\rightarrow\Psi e^{i\vartheta}
\end{equation}
with $\vartheta=\frac{1}{i}\lambda(x) e^{-i\omega t+iqy}$, gives rise to a spurious solution
\begin{equation}
\delta A_t=-\lambda\omega, \delta A_x=-i\frac{d\lambda}{dx}, \delta A_y=q\lambda, \delta\Phi=\lambda\Phi.
\end{equation}
This can be removed by requiring $\delta A_t=0$ at the AdS boundary. Furthermore, we require the last perturbation equation to be satisfied at the AdS boundary, namely
$\partial_z\delta A_t|_{z=0}=\frac{i}{\omega}(iq\partial_z\delta A_y+\partial_z\partial_x\delta A_x)|_{z=0}$. Regarding the other perturbed fields, we will impose the source free boundary condition at the AdS boundary. In addition, we impose Dirichlet boundary condition for $\delta\mathbf{A}$ and Neumann boundary condition for $\delta\Phi$  at $x=\pm\frac{R_x}{2}$. Then these boundary conditions, together with the above other perturbation equations, can be cast into the form $[\mathcal{A}(q)+\mathcal{B}(q)\omega]\delta= 0$ with $\delta$ the perturbation fields evaluated at the grid points by the pseudo-spectral method. Then the quasi-normal modes can be obtained numerically by solving this generalized eigenvalue problem. 
 \begin{figure}
\begin{center}
\includegraphics[width=8.0cm]{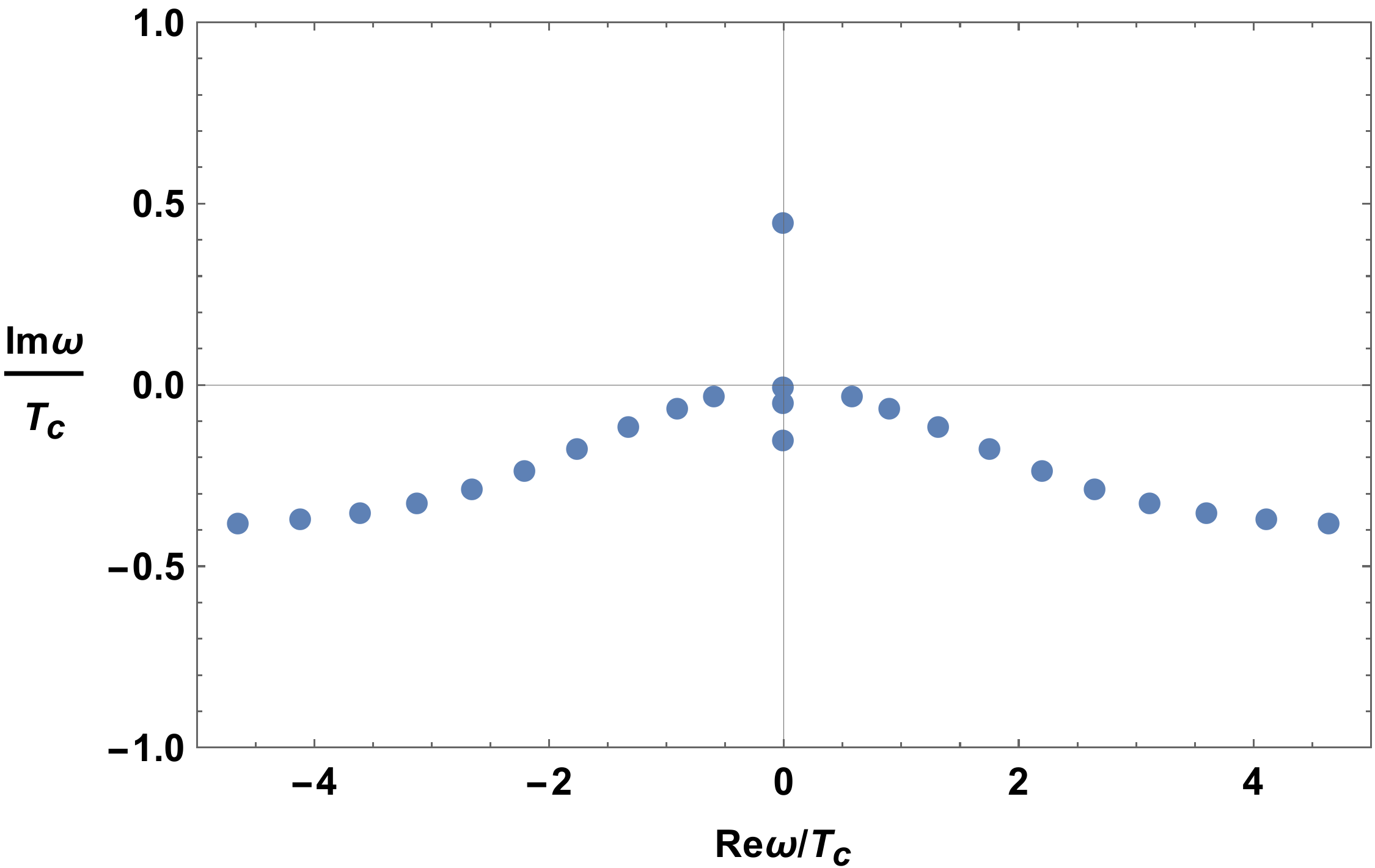}
\includegraphics[width=8.0cm]{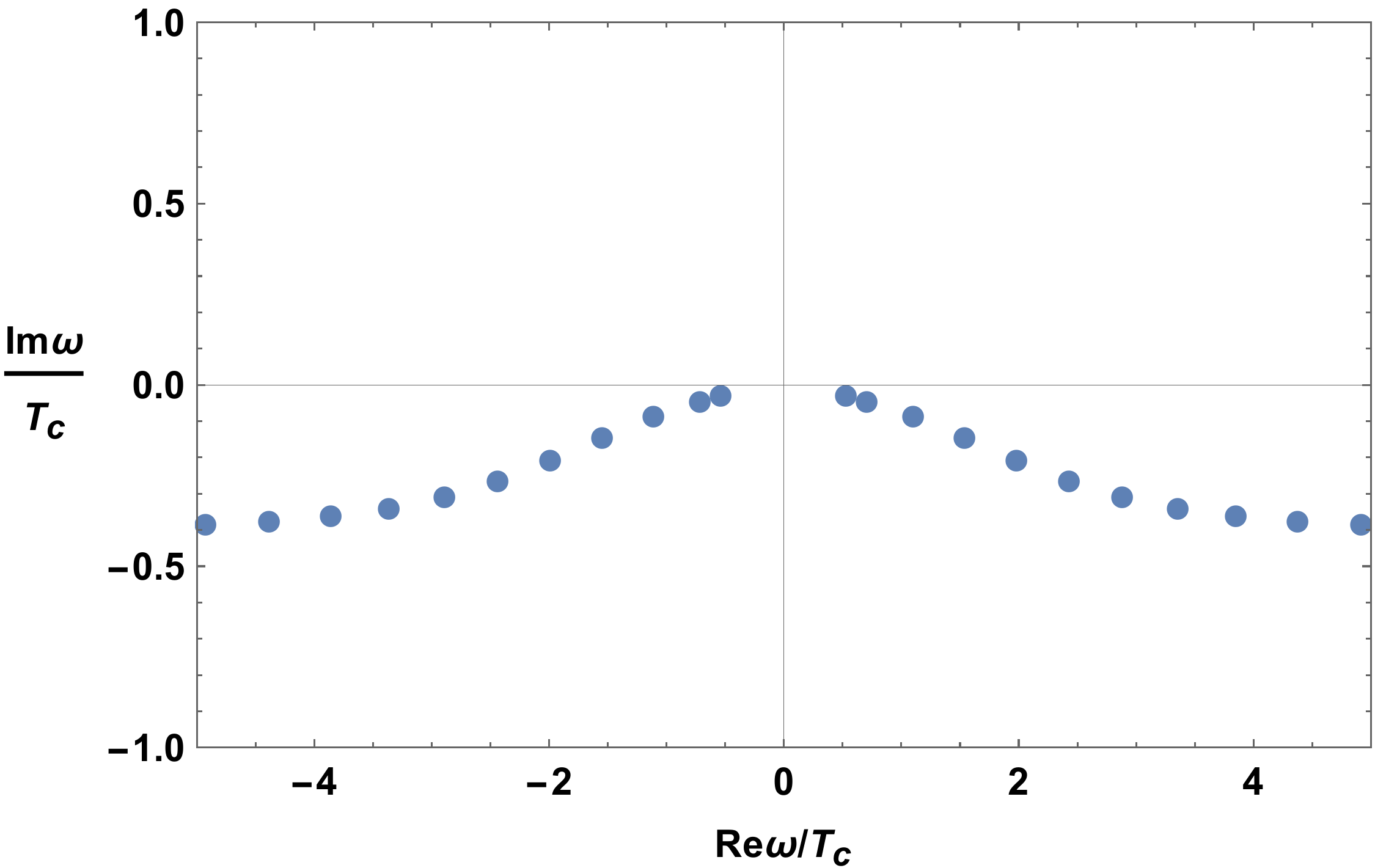}
\end{center}
\caption{The low lying quasi-normal modes with $qT_c=1$ on top of the dark soliton for the BCS-like superfluid at $\frac{T}{T_c}=0.78$. The upper panel is for the quasi-normal modes associated with $\delta\Phi$ of even parity. The lower panel is  for the quasi-normal modes associated with $\delta\Phi$ of odd parity.}
\label{qnmbcs}
\end{figure}

Note that the dark soliton background solution has even parity for $A_t$ and odd parity for $(\Phi, A_x)$ with respect to the $x$ axis. By inspection of the above linear perturbation equations on top of the dark soliton background, one can see that the quasi-normal modes with odd parity for $(\delta A_t,\delta A_y)$ and even parity for $(\delta\Phi,\delta A_x)$ decouple from the modes with opposite parity. So to reduce our computational resource for quasi-normal modes, we can restrict our computational domain onto $[0, \frac{R_x}{2}]$ in the $x$ direction by imposing Dirichlet or Neumann boundary condition at $x=0$, depending on whether the parity of the perturbed fields is odd or even. As a demonstration, we plot the spectrum of low lying quasi-normal modes in FIG.~\ref{qnmbcs}. As we can see, only the even parity branch of $\delta\Phi$ gives rise to the unstable mode, which is purely imaginary.

The strategy for the linear perturbation analysis of dissipative GPE is exactly the same, but much simpler. Here we present only the corresponding linear perturbation equation
\begin{eqnarray}
-i\omega(\delta a-\kappa \delta b)&=&(\frac{1}{2}q^2-\frac{1}{2}\frac{\partial^2}{\partial x^2}+a^2-\mu)\delta b,\\
-i\omega(-\delta b-\kappa \delta a)&=&(\frac{1}{2}q^2-\frac{1}{2}\frac{\partial^2}{\partial x^2}+3a^2-\mu)\delta a,\,\,\,\,\,\,\,
\end{eqnarray}
where we have also  decomposed the condensate wave function to its real and imaginary parts as $\psi=a+ib$ and assumed that the perturbation behaves like $\epsilon\delta(x)e^{-i\omega t+iqy}$.


\begin{thebibliography}{99}
\bibitem{large} For notational convenience we will set the overall factor $16\pi G q^2=1$ later on.
\bibitem{units}In what follows, we will work in the units where $L=1$ and $z_h=1$. The results presented later on can be obtained by the scaling symmetry of our theory.
\bibitem{periodic}In the fully nonlinear simulation, we impose the periodic boundary condition along the $y$ direction, so here $q=\frac{2\pi N}{R_y}$ with $N$ an integer and  $R_y$ the length along the $y$ direction.
\end{thebibliography}
\end{document}